\documentclass[prd,12pt,aps,showpacs,preprintnumbers,tightenlines,floatfix,nofootinbib]{revtex4}
\usepackage{epsfig}
\usepackage{amssymb}

\newcommand{\Z}{{Z \!\!\! Z}}
\newcommand{\beqn}{\begin{eqnarray}}
\newcommand{\eeqn}{\end{eqnarray}}
\newcommand{\eq}[1]{(\ref{#1})}
\newcommand{\const}{{\mathrm{const}}\,}
\newcommand{\dd}{\mathrm{d}}
\newcommand{\dD}{{\cal D}}
\newcommand{\dual}{\mbox{}^{\ast}}

\newcommand{\DFP}{\Delta_{FP}[\theta;\lambda]}
\newcommand{\DFPI}{\Delta^{-1}_{FP}[\theta;\lambda]}

\begin{document}

\title{Confinement and the photon propagator in $3D$ compact QED: \\
     a lattice study in Landau gauge at zero and finite temperature}

\author{M.~N.~Chernodub}
\affiliation{Institute of Theoretical and  Experimental Physics,
B. Cheremushkinskaja 25, Moscow, 117259, Russia}
\affiliation{Institute for Theoretical Physics, Kanazawa
University, Kanazawa 920-1192, Japan}
\author{E.--M.~Ilgenfritz}
\affiliation{Research Center for Nuclear Physics, Osaka University,
Osaka 567-0047, Japan}
\author{A.~Schiller}
\affiliation{Institut f\"ur Theoretische Physik and NTZ, Universit\"at
Leipzig, D-04109 Leipzig, Germany}

\preprint{LU-ITP 2002/012}
\preprint{KANAZAWA-02-17}
\preprint{ITEP-LAT-2002-08}
\preprint{RCNP-Th02013}

\begin{abstract}
On the lattice we study the gauge boson propagator of three
dimensional compact QED in Landau gauge at zero and non-zero temperature.
The non-perturbative effects are taken into account by the generation
of a mass, by an anomalous dimension and by the photon wave function
renormalization. All these effects can be attributed
to the monopoles: they are absent in the propagator of the
singularity-free part of the gauge field.
We assess carefully the Gribov copy problem for the propagator and
the parameters emerging from the fits.
\end{abstract}


\date{\today}

\maketitle

\section{Introduction}
\label{sec:introduction}

Three--dimensional compact electrodynamics (cQED$_3$) has two
essential features in common with QCD, confinement~\cite{Polyakov}
and chiral symmetry breaking~\cite{ChSB}. Although the physics
behind it might be very different, monopole dynamics in three and
in four dimensions, it is amusing to study certain
nonperturbative aspects within a lower--dimensional model
such as cQED$_3$. Apart from its role as a toy model for QCD, the
nonperturbative properties of cQED$_3$ deserve attention in
themselves because this model was shown to describe some features
of Josephson junctions~\cite{Josephson} and high--$T_c$
superconductors~\cite{HighTc}. cQED$_3$ was the first
example, in space--time dimensions greater than two where it
becomes a nontrivial problem, in which confinement of
electrically charged particles was understood
analytically~\cite{Polyakov}. It is the result of the dynamics of
monopoles which emerge due to the compactness of the gauge field.
Other common features of cQED$_3$ and QCD are the existence of a
mass gap and of a confinement--deconfinement phase transition at
some nonzero temperature.

In two recent papers we have demonstrated how the deconfinement
phase transition, occurring under the influence of
compactification of one dimension, in $2+1$ dimensions, finds its
explanation from the monopole point of view~\cite{CISPaper1} and
why the deconfinement phase transition is independent on the
strength of the external fields~\cite{CISPaper2}.

In a recent letter~\cite{CISLetter} we answered the question what
effect the confinement property has on the gauge boson propagator
in this theory and what of the propagator is changing at the
deconfinement temperature. The effects are twofold: first, an
anomalous dimension appears which modifies the momentum
dependence, and second, a mass is generated which can be well
understood in terms of Polyakov's theory~\cite{Polyakov}. In
Ref.~\cite{CISLetter} we also have used the unique possibility in
lattice simulations to remove the monopole degrees of freedom from
the quantum gauge field in order to show that all nontrivial
effects reside exclusively in the singular fields of the
monopoles.

Note that a non--trivial anomalous dimension of the gauge boson
propagator may also appear due to dynamical matter
fields~\cite{AnomalousMatter}. The sign (in our notation) of the
anomalous dimension induced in this way is different and this
leads to the binding of the monopoles into dipole pairs and,
consequently, to the disappearance of the confinement at certain
distances between the test particles. This picture was confirmed
for the compact model in the presence of the dynamical matter
fields~\cite{MatterFields}.

In the present paper we want to extend the analysis of
Ref.~\cite{CISLetter} in various respects. Since the propagator is
studied in the Landau gauge, first of all we have investigated
more carefully the quality of the gauge fixing and the importance
of the Gribov copy problem, first on the propagator itself and
then on the parameters that finally describe the functional form
of the propagator.

In a second direction, we investigate the potential influence that
different definitions of the gauge potential $A_{\mu}$ in terms of
the lattice link fields might have on the resulting propagator.
One of the choices is strongly recommended by the explicit gauge
invariance (transversality) of the emerging propagator. However,
in the case when the longitudinal part does not vanish by
construction, proper selection of the transverse component also leads
to almost the same fit parameters. The coincidence becomes
obviously better at larger $\beta$.

Going over from zero to finite temperature, it is important to
realize that more structure functions are necessary to the
describe the finite-temperature case. One of them, the
deconfinement-sensitive $D_L$, was studied already in
Ref.~\cite{CISLetter}, while the other, $D_T$, was found to
be extremely sensitive to the Gribov problem. Only exerting extreme
care can one expose the change going on at the deconfinement
transition.

The paper is organized as follows. In Sect.
\ref{sec:model_and_propagator} we define the lattice model and the
tensorial structure of the propagators, at $T = 0$ and for $T \ne
0$. Sect. \ref{sec:analytical_consideration} contains a
discussion of the minimal Landau gauge which points out the
topological aspects related to the gauge fixing, in particular the
condition of having a minimal total length of Dirac strings. In
Sect. \ref{sec:algorithms} we report on the numerical algorithms
for updating and gauge fixing which we have used in our
investigation. The following Sect. \ref{sec:zero_temperature} is
devoted to an outline the results for $T = 0$. Sect.
\ref{sec:finite_temperature} describes the particular requirements
of gauge fixing in the $T \ne 0$ case and contains our results for
the propagators $D_L$ and $D_T$. Some conclusions are formulated
in Sect. \ref{sec:conclusions}.

\section{The $3D$ compact U(1) model and the photon propagator}
\label{sec:model_and_propagator}

For the compact U(1) model we chose the Wilson single-plaquette action:
\beqn
S[\theta_l] = \beta \sum\limits_{p} \big( 1 - \cos \theta_p \big) \, ,
\label{def:action}
\eeqn
where $\theta_p$ is the $U(1)$ field strength tensor represented
by the plaquette curl of the compact link field $\theta_l$.
The lattice is three-dimensional, and the
basic degrees of freedom are the links $U_l=\exp\left(i~\theta_l \right)$.
The measure of the link angles is flat over the interval
$-\pi < \theta \leq \pi$.
The lattice coupling constant $\beta$ is related to the lattice
spacing $a$ and the continuum coupling constant $g_3$ of the $3D$
theory,
\beqn
\beta = 1 \slash (a\, g^2_3)\; .
\label{def:beta}
\eeqn
Note that in three dimensional gauge theory the coupling constant $g_3$ has
dimension ${\mathrm{mass}}^{1 \slash 2}$.
Zero physical temperature is represented by symmetric lattices,
$L_t=L_s$.~\footnote{$L_t$ is the extension in the third ($z$) direction.}
The lattice corresponding to finite temperature is asymmetric, $L^2_s\times
L_t$, $L_t << L_s$. In the limit $L_s \to \infty$, the temporal extension
of the lattice is related to the physical temperature,
$L_t = 1 \slash (T a)$.
Using (\ref{def:beta}) the temperature is given in units of $g_3^2$
in terms of the lattice parameters as follows:
\beqn
\frac{T}{g^2_3} = \frac{\beta}{L_t} \, .
\label{def:temperature}
\eeqn
Our simulations for zero temperature were performed mainly on
a $32^3$ lattice, those at finite temperature on a $32^2 \times 8$ lattice.

The final discussion of the photon propagator will be given in lattice
momentum space.
Being always defined in a specified gauge, the propagator is written in
terms of the Fourier transformed gauge potential,
\beqn
\tilde{A}_{{\vec k},\mu} = \frac{1}{\sqrt{L_1~L_2~L_3}}
\sum\limits_{{\vec n}}
\exp \Bigl( 2 \pi i~\sum_{\nu=1}^{3} \frac{k_{\nu}
(~n_{\nu}+\frac{1}{2}\delta_{\nu\mu}~) } {L_{\nu}} \Bigr)
~A_{{\vec n}+\frac{1}{2}{\vec \mu},\mu} \; ,
\label{def:fourier_transformation}
\eeqn
which is a sum over a certain discrete set of points
${\vec x}={\vec n}+\frac{1}{2}{\vec \mu}$ forming the support of
$A_{{\vec x},\mu}$ on the lattice.
These are the midpoints of the links in $\mu$ direction while
${\vec n}$ denotes the lattice sites (nodes) with integer Cartesian
coordinates.
The propagator is the gauge-fixed ensemble average
of the following bilinear in $\tilde{A}$,
\beqn
D_{\mu\nu}({\vec p}) = \langle \tilde{A}_{ {\vec k},\mu}
                               \tilde{A}_{-{\vec k},\nu} \rangle \; .
\label{def:propagator}
\eeqn
Two identifications of $A_{{\vec x},\mu}$ were adopted in the literature
and will be compared in our paper: the {\it angle}--definition
\beqn
A_{{\vec n}+\frac{1}{2}{\vec \mu},\mu}
    = \theta_{{\vec n},\mu}/(g_3~a)
    = \log \left( U_{{\vec n},\mu} \right)/(i~g_3~a)
\label{def:angle}
\eeqn
and the {\it sine}--definition
\beqn
A_{{\vec n}+\frac{1}{2}{\vec \mu},\mu}
    = \sin \left( \theta_{{\vec n},\mu} \right)/(g_3~a)
    = \left( U_{{\vec n},\mu} - U^{*}_{{\vec n},\mu} \right)/(2~i~g_3~a) \, .
\label{def:sine}
\eeqn
The corresponding propagators will be denoted as
$D^{ang}_{\mu\nu}$ or $D^{sin}_{\mu\nu}$, respectively.

The lattice momenta ${\vec p}$ on the left hand side of (\ref{def:propagator})
are related to the integer valued Fourier momenta ${\vec k}$ as follows:
\beqn
p_\mu(k_\mu)=  \frac{2}{a} \sin \frac{ \pi k_\mu}{L_\mu}
\,, \quad  k_\mu=0, \pm 1,..., \pm \frac{L_\mu}{2} \; .
\label{def:momenta}
\eeqn
The lattice equivalent of $p^2 = {\vec p}^2$ is in $3D$
\beqn
p^2({\vec k})=
\frac{4}{a^2} \sum_{\mu=1}^3 \left(\sin \frac{\pi k_\mu}{L_\mu}
\right)^2  \, .
\label{def:momentum_squared}
\eeqn

At zero temperature, based on Euclidean rotational invariance,
the continuum propagator would be expressible by functions of $p^2$.
The most general tensor structure is then the following one including
two scalar functions of $p^2$,
\beqn
D_{\mu\nu}({\vec p})=
P_{\mu\nu}({\vec p})~D(p^2) + \frac{p_\mu p_\nu}{p^2} \frac{F(p^2)}{p^2}
\label{def:tensor_structure}
\eeqn
with the $d$-dimensional (in our case $d=3$) transverse projection operator
\beqn
P_{\mu\nu}({\vec p}) = \delta_{\mu\nu}- \frac{p_\mu p_\nu}{p^2} \, .
\label{def:transverse_proj}
\eeqn
The projector has the properties
\beqn
P_{\mu\alpha}({\vec p})~P_{\alpha\nu}({\vec p}) = P_{\mu\nu}({\vec p}) \,,
\quad    P_{\mu\mu}({\vec p}) = d-1
 \, .
\eeqn
The two structure functions $D(p^2)$ and $F(p^2)$
can be extracted by projection, on the lattice from $D_{\mu\nu}({\vec p})$
according to (\ref{def:propagator}), as
\beqn
F(p^2)=  \sum\limits_{\mu,\nu=1}^{3} p_\mu~ D_{\mu\nu}({\vec p})~ p_\nu
\label{def:F_project_1}
\eeqn
and
\beqn
p^2~D(p^2)= \frac{1}{d - 1}~P_{\mu\nu}({\vec p})~D_{\mu\nu}({\vec p}) \, .
\eeqn
They are found, in the best case, only approximately rotationally
invariant, {\it i.e.} individual momenta ${\vec p}$ might slightly differ
in the function values $D$ or $F$ they provide, even if they have the
same $p^2$.
Dense data points  close together in $p^2$ might scatter rather
than forming a smooth function of $p^2$.

In practice, using these definitions, we extract at first the function
$F(p^2)$, in the $d=3$ case through
\begin{eqnarray}
F(p^2) & = & p_1^2 D_{11}({\vec p})
           + p_2^2 D_{22}({\vec p})
           + p_3^2 D_{33}({\vec p}) \nonumber \\
& + & 2 p_1 p_2 \mathrm {Re} D_{12}({\vec p})
    + 2 p_1 p_3 \mathrm {Re} D_{13}({\vec p})
    + 2 p_2 p_3 \mathrm {Re} D_{23}({\vec p}) \, .
\label{def:F_project_2}
\end{eqnarray}
The imaginary parts of non-diagonal $D_{\mu\nu}$ cancel in the sum and
were omitted. Then, the function $D(p^2)$ is obtained through
\beqn
D(p^2)= \frac{1}{d-1}\left[\left( D_{11}({\vec p})
                                + D_{22}({\vec p})
                                + D_{33}({\vec p}) \right)
           - F(p^2)/p^2 \right] \, .
\label{def:D_project_2}
\eeqn

If the Landau gauge would be exactly fulfilled, one would
expect that  $F(p^2) \equiv 0$.
On the lattice, in the case of the {\it sine}--definition
for $A_{{\vec x},\mu}$, this is actually the case as soon as
one of the Gribov copies is reached, with an accuracy which directly
reflects the stopping precision of the gauge fixing procedure
(as will be discussed below).
In this case, a simplified definition in terms of
the diagonal components $D_{\mu\mu}({\vec p})$ would be appropriate.
In case of  the {\it angle}--definition the structure function $F(p^2)$ does
not vanish, therefore all components of $D_{\mu\nu}({\vec p})$ contribute to
$D(p^2)$.

For the finite temperature case, the propagator lacks $O(3)$ rotational
symmetry. Now we have to consider two scalar functions, $D_T$ and $D_L$
instead of $D$, each multiplying the $(d-1)$-dimensional transverse
projection operator $P^T$ and the $(d-1)$-dimensional longitudinal projection
operator $P^L$, respectively. The scalar functions $D_T$, $D_L$ and $F$
depend now separately on the length of the space--like part of ${\vec p}$
with $i=1,\dots,d-1$,
\beqn
{\mathbf p}^2 = p_1^2 + \dots + p_{d-1}^2 \, , \quad
|{\mathbf p}|=\sqrt{{\mathbf p}^2} \, ,
\label{def:space_length}
\eeqn
and the ``temporal'' momentum $p_d$:
\beqn
D_{\mu\nu}({\vec p})= P^T_{\mu\nu}({\vec p}) D_T(|{\mathbf p}|,p_d)
                   + P^L_{\mu\nu}({\vec p}) D_L(|{\mathbf p}|,p_d)
              + \frac{p_\mu p_\nu}{p^2} \frac{F(|{\mathbf p}|,p_d)}{p^2} \, .
\label{def:all_components_at_T}
\eeqn
The two projection operators are defined as follows ($i,j=1,\dots,d-1$):
a transverse one,
\beqn
P^T_{ij}({\vec p}) = \delta_{ij}- \frac{p_i p_j}{{\mathbf p}^2} \,, \quad
P^T_{dd}({\vec p}) = P^T_{di}({\vec p}) = P^T_{id}({\vec p}) = 0 \, ,
\label{def:PT}
\eeqn
and a longitudinal one
\beqn
P^L_{\mu\nu}({\vec p}) = P_{\mu\nu}({\vec p}) - P^T_{\mu\nu}({\vec p}) \, .
\label{def:PL}
\eeqn
Obviously, these projectors have the properties
\begin{eqnarray}
P^T_{\mu\alpha}({\vec p})~P^T_{\alpha\nu}({\vec p}) & = &
P^T_{\mu\nu}({\vec p}) \, , \quad  P^T_{\mu\mu}({\vec p}) = d-2 \\
P^L_{\mu\alpha}({\vec p})~P^L_{\alpha\nu}({\vec p}) & = &
P^L_{\mu\nu}({\vec p}) \, , \quad  P^L_{\mu\mu}({\vec p}) = 1 \\
P^L_{\mu\alpha}({\vec p})~P^T_{\alpha\nu}({\vec p}) & = & 0  \, .
\label{def:projector_properties}
\end{eqnarray}
The scalar functions $D_T$ and $D_L$ can be extracted from
$D_{\mu\nu}({\vec p})$ via
\begin{eqnarray}
P^T_{\mu\nu}({\vec p})~D_{\mu\nu}({\vec p}) &
= & (d-2)~D_T(|{\mathbf p}|,p_d) \\
P^L_{\mu\nu}({\vec p})~D_{\mu\nu}({\vec p}) &
= & D_L(|{\mathbf p}|,p_d) \nonumber \\
& = & (d-1)~D(|{\mathbf p}|,p_d) - (d-2)~D_T(|{\mathbf p}|,p_d) \,.
\label{def:DT_and_DL}
\end{eqnarray}
For $d=3$ we can write down explicitly the definitions
\beqn
{\mathbf p}^2 D_T(|{\mathbf p}|,p_d) =
            p_2^2 D_{11}({\vec p})
                 + p_1^2 D_{22}({\vec p})
- 2 p_1 p_2~\mathrm {Re} D_{12}({\vec p})
\label{def:DT_definition}
\eeqn
and
\begin{eqnarray}
{\mathbf p}^2 ( {\mathbf p}^2 + p_3^2 ) D_L(|{\mathbf p}|,p_d) &=&
{\mathbf p}^2 \left( {\mathbf p}^2 D_{33}({\vec p})
         - 2 p_1 p_3 {\mathrm {Re}}D_{13}({\vec p})
         - 2 p_2 p_3 {\mathrm {Re}}D_{23}({\vec p}) \right)
\nonumber \\
&+&
       p_3^2 \left( p_1^2 D_{11}({\vec p})
                  + p_2^2 D_{22}({\vec p})
+ 2 p_1 p_2 {\mathrm {Re}}D_{12}({\vec p}) \right) \, .
\label{def:DL_definition}
\end{eqnarray}
In the static limit, $p_3=0$,
\beqn
D_L(|{\mathbf p}|,p_3=0) \equiv D_{33}(|{\mathbf p}|,p_3=0) \, .
\label{def:static_limit}
\eeqn
This propagator for the case of the {\it angle}--definition and its fit in
terms of mass and anomalous dimension as well as the changes at the
deconfinement transition were discussed in Ref.~\cite{CISLetter}.

In the case of $T=0$, the data to be presented below will be
averaged over measurements of these quantities obtained for
different ${\vec k}=(k_1,k_2,k_3)$
giving rise to the {\it same} $p^2$ according to (\ref{def:momentum_squared}).
In the case $T \ne 0$ we will show data averaged over
different $(k_1,k_2)$ giving rise to the {\it same} ${\mathbf p}^2$
according to
(\ref{def:space_length}). One should keep in mind that by this ``trick'' one
enforces the rotational invariance ``by hand'', and the statistical errors
are reduced as compared to measurements for individual ${\vec k}$-components
of the propagator.

In order to discuss the functional form of the propagator
from the viewpoint
of confinement effects, with confinement being induced
by the monopole plasma,
we will decompose the gauge fields into singular (monopole)
and regular (photon) contributions on the level of the
link angles,
\beqn
\theta_{{\vec n},\mu} = \theta_{{\vec n},\mu}^{phot}
+ \theta_{{\vec n},\mu}^{mono} \, ,
\label{def:theta_decomposition}
\eeqn
by a procedure to be described below. After the decomposition
(\ref{def:theta_decomposition}) of the link angles is done,
one may define the corresponding
$A^{phot}_{{\vec n}+\frac{1}{2}{\vec \mu},\mu}$ and
$A^{mono}_{{\vec n}+\frac{1}{2}{\vec \mu},\mu}$ through the
{\it angle}--definition (\ref{def:angle}) or the
{\it sine}--definition (\ref{def:sine}), respectively.
Then, by fast Fourier transform, the corresponding
$\tilde{A}^{phot}_{{\vec k},\mu}$ and
$\tilde{A}^{mono}_{{\vec k},\mu}$ are evaluated.
For each configuration and a certain set of momenta
the bilinears for the photon part
$\tilde{A}^{phot}_{ {\vec k},\mu} \tilde{A}^{phot}_{-{\vec k},\nu}$, the
monopole part
$\tilde{A}^{mono}_{ {\vec k},\mu} \tilde{A}^{mono}_{-{\vec k},\nu}$ and
the mixed bilinear
$\tilde{A}^{phot}_{ {\vec k},\mu} \tilde{A}^{mono}_{-{\vec k},\nu}$
are formed. These are the observables which are associated -- by averaging
over the Monte Carlo ensemble --
to the propagators
$D^{phot}_{\mu\nu}({\vec p})$ (the photon or {\it regular} propagator),
$D^{mono}_{\mu\nu}({\vec p})$ (the {\it singular} propagator)
and
$D^{mixed}_{\mu\nu}({\vec p})$ (the {\it mixed} propagator).
These propagators are considered together with the {\it full} propagator
which uses the original link angles $\theta_{{\vec n},\mu}$ before the
splitting (\ref{def:theta_decomposition}) was performed.

In the $T=0$ case, all these functions of ${\vec p}$ are then mapped by the
projections (\ref{def:F_project_2},\ref{def:D_project_2}) to the scalar
structure functions
$D^{phot}(p^2)$, $D^{mono}(p^2)$, $D^{mixed}(p^2)$ and
$F^{phot}(p^2)$, $F^{mono}(p^2)$, $F^{mixed}(p^2)$.
In the finite temperature case we proceed analogously.

For the {\it angle}--definition of the vector potential we stress
that the structure functions obviously satisfy exact additivity
\beqn
D({\vec p}) & = & D^{phot}({\vec p}) + D^{mono}({\vec p})
+ 2~D^{mixed}({\vec p})    \\
F({\vec p}) & = & F^{phot}({\vec p}) + F^{mono}({\vec p})
+ 2~F^{mixed}({\vec p}) \, .
\label{def:D_F_decomposition}
\eeqn

\section{Monopoles, Dirac strings, and the minimal Landau gauge}
\label{sec:analytical_consideration}

In this section we partially follow Ref.~\cite{PoZuYe}.
Gauge fixing to the Landau gauge means, for a given
configuration $\theta_l=\theta_{{\vec n},\mu}$,
to find a gauge transformations $\omega_{\vec n}$ such
that, for a {\it gauge functional}
\beqn
{\cal F}(\theta)=\sum\limits_l \cos(\theta_l)
\label{def:cos_gauge_functional}
\eeqn
the transformed gauge functional becomes maximal,
\beqn
\max\limits_{\omega} {\cal G}(\theta,\omega) \, ,
\quad {\cal G}(\theta,\omega) = {\cal F}(\theta^{(\omega)}) \, .
\label{def:cos_gauge_condition}
\eeqn
Here $\theta_{l}^{(\omega)}$ denotes the gauge transformed gauge
field~\footnote{Here and in the following we use the
differential form notation on the lattice:  $(a,b) = \sum_l a_l\,
b_l$, ${||\theta||}^2 = (\theta,\theta)$. The operations
$\dd \theta$ and $\delta \theta$ are the lattice curl and divergence,
respectively. The Laplacian is denoted as $\Delta = \delta \dd +
\dd \delta$.}
\beqn
\theta \to \theta^{(\omega)} = {\theta + \dd \omega}_{2\pi}
\equiv \theta + \dd \omega + 2 \pi k\,,\quad k \in \Z \, ,
\label{def:form_gauge_transformation}
\eeqn
where the integer number $k = k(\theta,\omega)$ for each link is chosen such
that $\theta^{(\omega)} \in (-\pi,\pi]$.

Instead of (\ref{def:cos_gauge_condition}),
following Ref.~\cite{PoZuYe} we use for the purpose of this section
in the following the Villain form of the gauge condition
\beqn
\min\limits_{\omega} {||\theta^{(\omega)}||}^2 \, .
\label{def:villain_gauge_condition}
\eeqn

The Faddeev--Popov (FP) determinant is introduced by
the following decomposition of unity
\beqn
1 = \DFP ~ \int\limits^\pi_{- \pi} \dD \omega \,
e^{- \lambda {||\theta^{(\omega)}||}^2} \, ,
\label{def:faddeev_popov_1}
\eeqn
where $\lambda$ is the gauge fixing parameter.
In order to achieve the gauge (\ref{def:villain_gauge_condition})
we have to send $\lambda$ to infinity. In this case the limits of integration
in (\ref{def:faddeev_popov_1})
can be extended to $\pm \infty$ since the saddle point
approximation is exact in the limit $\lambda \to \infty$. Moreover, the
integer valued variable $k$ in (\ref{def:form_gauge_transformation})
becomes effectively independent on
$\theta$ and $\omega$ since the values of $k$ for which
$\theta^{(\omega)} \notin (-\pi,\pi]$ are exponentially suppressed.
Therefore, in the limit of infinite $\lambda$, the FP determinant
(\ref{def:faddeev_popov_1}) can be written as follows:
\beqn
\DFPI = \int\limits^\infty_{- \infty} \dD \omega \,
\sum\limits_{k \in \Z(c_1)}\,
\exp\Bigl\{- \lambda {||\theta + \dd \omega + 2 \pi k||}^2 \Bigr\} \, .
\label{def:faddeev_popov_2}
\eeqn

Using the Hodge--de-Rahm transformation, $k = \delta \Delta^{-1} \dd  k
+ \dd \Delta^{-1} \delta k$, and making the shift $\omega \to
\omega - \Delta^{-1} \delta k$ we get
\beqn
\DFPI = \const \int\limits^\infty_{- \infty} \dD \omega \,
\sum\limits_{\stackrel{s \in \Z(c_2)}{\dd s = 0}}\,
\exp\Bigl\{- \lambda {||\theta + \dd \omega + 2 \pi \delta
\Delta^{-1} s||}^2 \Bigr\} \, ,
\label{def:faddeev_popov_3}
\eeqn
where we have changed the variables, $s = \dd k$. The integration over
$\omega$ gives
\beqn
\DFPI = \const \sum\limits_{\stackrel{s \in \Z(c_2)}{\dd s = 0}}\,
\exp\Bigl\{- \lambda \Bigl(\dd\theta + 2 \pi \delta \Delta^{-1} s,
\Delta^{-1} (\dd\theta + 2 \pi \delta \Delta^{-1} s \Bigl) \Bigr\} \, .
\label{def:faddeev_popov_4}
\eeqn

To proceed further we separate the gauge field $\theta$
into regular (photon) and singular (monopole)
parts following Ref.~\cite{PhMon},
\beqn
\theta = \theta^{phot} + \theta^{mono}\,, \quad
\theta^{mono} = 2 \pi \Delta^{-1} \delta p[j]\,,
\label{def:form_theta_decomposition_1}
\eeqn
where the dual one-form $\dual j$ represents the monopoles on the
dual lattice sites. The one--form on the dual
lattice, $p[j]$, defines the Dirac lines
that connect the monopoles and
anti--monopoles, $\delta \dual p[j] = \dual j$.

The photon part $\theta^{phot}$ is free of singularities
while the monopole part $\theta^{mono}$ contains the information about
all monopole singularities:
\beqn
\frac{1}{2\pi} \dd {[\dd \theta^{phot}]}_{2\pi} = 0 \,, \quad
\frac{1}{2\pi} \dd {[\dd \theta^{mono}]}_{2\pi} = j \, .
\label{def:form_theta_decomposition_2}
\eeqn
Here the DeGrand--Toussaint definition of the monopole~\cite{DGT} was used.
Substituting (\ref{def:form_theta_decomposition_1}) into
(\ref{def:faddeev_popov_4}) we get, after a little algebra,
\beqn
\DFPI & = & \const \exp \Bigl\{ 4 \lambda (j, \Delta^{-2} j)\Bigr\} \,
\sum\limits_{\stackrel{s \in \Z(c_2)}{\dd s = 0}}\,
\exp \Bigl\{- \lambda S_{gf}(\theta^{phot},p[j]+s)\Bigr\} \, ,
\label{def:faddeev_popov_5}
\eeqn
with
\beqn
S_{gf}(\theta^{phot},p) & = & \Bigl(\dd\theta^{phot} + 2 \pi p,
\Delta^{-1}
(\dd\theta^{phot} + 2 \pi (p[j] + s)) \Bigr) \, .
\label{def:s_gauge_fix}
\eeqn
The meaning of the last equations is the following.
The gauge transformation
(\ref{def:form_gauge_transformation}) contains both regular
($\dd \omega$) and singular ($k$) parts.
The former
transforms the photon part of the gauge field while the latter
changes the monopole part shifting the Dirac string (but leaving the
monopoles $j$ intact). We have already integrated out the regular gauge
transformations, therefore (\ref{def:faddeev_popov_5})
depends explicitly on $\dd \theta^{phot}$ which is invariant under
regular gauge transformations,
$\theta^{phot} \to \theta^{phot} + \dd \omega$.
The sum in (\ref{def:faddeev_popov_5})
over all possible shifts of the Dirac lines,
\beqn
\dual p[j] \to \dual p[j] + \dual s \, ,
\label{def:shifts}
\eeqn
corresponds to the integration over all singular gauge transformations
(remember that $\dual s$ is the closed line on the dual lattice,
$\delta \dual s = 0$). Thus \eq{def:faddeev_popov_5}
is implicitly invariant under
the singular gauge transformations as well.

In the limit $\lambda \to \infty$ the only contribution to the FP determinant
is given by the {\it global} minimum of the gauge fixing functional
(\ref{def:s_gauge_fix}) with respect
to the variations (\ref{def:shifts}) of the Dirac line,
\beqn
S^{min}_{gf}(\theta^{phot},j) = \min\limits_{\dd s = 0}
S_{gf}(\theta^{phot},p[j]+s) \, .
\label{def:minimum}
\eeqn
If the photon field is absent, the minimum (\ref{def:minimum})
is given by the Dirac line with minimal ``Coulomb interaction'' ({\it c.f.}
(\ref{def:s_gauge_fix}) ).
For a lattice monopole and anti--monopole separated along
one axis this line is the shortest path connecting the pair.

We substitute the FP unity
(\ref{def:faddeev_popov_1},\ref{def:faddeev_popov_5}) into the partition
function of compact electrodynamics,
\beqn
Z = \int\limits^\pi_{-\pi} \dD \theta \, e^{- S(\theta)} \, ;
\label{def:partition_function_1}
\eeqn
then we transform the gauge field, $\theta \to \theta^{(- \omega)}$ and
get the product of the gauge orbit volume, $\int \dD \omega$, and the
partition function within the fixed gauge,
\beqn
Z_{gf} = \int\limits^\pi_{-\pi} \dD \theta \, e^{- S(\theta)
- \lambda {||\theta||}^2} \, \DFP \, .
\label{def:partition_function_2}
\eeqn

Separating the gauge field on the monopole and photon parts
as indicated by (\ref{def:form_theta_decomposition_1})
and using the Hodge--de-Rahm transformation, one can show that
\beqn
{||\theta||}^2 = (\delta \theta^{phot}, \Delta^{-1} \delta \theta^{phot})
+ \Bigl(\dd \theta^{phot} + 2 \pi p[j], \Delta^{-1}
(\dd \theta^{phot} + 2 \pi p[j]) \Bigr) - 4 \pi^2 (j,\Delta^{-2} j) \, .
\label{def:min_norm_squared}
\eeqn
According to eqs.
(\ref{def:faddeev_popov_5},\ref{def:minimum},\ref{def:partition_function_2})
the only non--vanishing contribution to the partition function in the limit
$\lambda \to \infty$ comes from the global minimum
(\ref{def:min_norm_squared})
in the gauge orbit. Comparison of (\ref{def:minimum}) and
(\ref{def:min_norm_squared}) shows
that this minimum is defined by the following conditions:
\beqn
\delta \theta^{phot}  =  0
\label{def:condition_1}
\eeqn
together with
\beqn
S_{gf}(\theta^{phot},p[j])  =  S^{min}_{gf}(\theta^{phot},p[j]) \, ,
\label{def:condition_2}
\eeqn
where $S_{gf}$ and $S^{min}_{gf}$ are given in (\ref{def:s_gauge_fix})
and (\ref{def:minimum}), respectively.

In the continuum limit the condition (\ref{def:condition_1})
leads to the usual Landau gauge condition,
\beqn
\partial_\mu A^{phot}_\mu = 0 \, ,
\label{def:continuum_condition}
\eeqn
while the condition (\ref{def:condition_2})
can be formulated as a requirement for
the Dirac lines to form a configuration with as small as possible length:
\beqn
\min\limits_{\delta \dual p[j] = \delta j} \,
{\mathrm{length}}\Bigl(p[j]\Bigr) \,.
\label{def:minimal_length}
\eeqn
Indeed, the Dirac lines, $\dual p[j] = \dual p_1[j] + \dual p_2[j]
+ \dots$ , correspond to the singular $\delta$--functions in the
continuum limit. Here $\dual p_i$ correspond to mutually unconnected
pieces of these lines. The self--interaction of the Dirac
lines in (\ref{def:condition_2}),
$\sum_i (\dual p_i[j],\Delta^{-1} \dual p_i[j])$,
contains the term $\alpha \sum_i {\mathrm{length}}\Bigl(p_i[j]\Bigr)$
(with a logarithmically divergent coefficient $\alpha$) plus finite terms.
The ``Coulomb interaction'' of different pieces of the Dirac lines,
$(\dual p_i[j],\Delta^{-1} \dual p_k[j])$, $i \neq k$, as well as
the contribution of the regular fields into the condition
(\ref{def:condition_2}) are finite in the continuum limit.
Thus the only essential contribution to the
condition (\ref{def:condition_2}) in the continuum limit is given
by the term
$\alpha \sum_i {\mathrm{length}}\Bigl(p_i[j]\Bigr) \equiv
\alpha~{\mathrm{length}}\Bigl(p[j]\Bigr)$, which gives the
condition~(\ref{def:minimal_length}).

Thus we conclude that in the continuum limit the minimal Landau
gauge for the compact gauge fields is reduced to the local gauge
condition (\ref{def:continuum_condition}) for the regular fields
and a non--local condition (\ref{def:minimal_length}) -- the
requirement for the total length of the Dirac lines to be as small
as possible -- for the singular fields. This result can easily be
generalized to the $4D$ case.

\section{Numerical algorithms used in the analysis}
\label{sec:algorithms}

\subsection{Monte Carlo Updating}
\label{subsec:MC}

The Monte Carlo algorithm in use for this investigation is a mixture of
local and global
updates. The local Monte Carlo algorithm is based on a 5--hit Metropolis
update sweep in an even-odd fashion, alternating with a microcanonical
sweep, also in checkerboard mode. Both together are considered as {\it one}
local update. After three local updates the Metropolis step width is
eventually tuned to keep an acceptance in the range between 40 \% and 60 \%.

For better ergodicity, in particular in the presence of an external field
(considered in Ref. \cite{CISPaper2}), global updates
have also been included,
following the ideas of Ref.~\cite{DamgaardHeller}.
In the equilibrium regime, after every three complete local updates, a global
refreshment step is attempted.
We try to add one unit of flux to the dynamical gauge field,
with random sign in one of the three directions randomly selected.
The proposed flux addition is subject to a global Metropolis acceptance check.

For example, one unit of flux in the $\mu\nu$ plane is introduced with the
help of the following gauge field shift~\cite{DamgaardHeller}
$ \theta_{{\vec x},\mu} \rightarrow
[ \theta_{{\vec x},\mu}
           + \tilde{\theta}_{{\vec x},\mu} ]_{ {\mathrm{mod}}~2\pi } $:
\beqn
\tilde{\theta}_{{\vec x},\nu} & = & \frac{\pi}{L_{\mu}}
(2~x_{\mu} - L_{\mu} - 1) \,, \quad
\tilde{\theta}_{{\vec x},\nu} = 0 \,\,\,\, {\mathrm{for}} \,\,\,\,
x_{\nu} \neq L_{\nu} \,, \nonumber \\
\tilde{\theta}_{{\vec x},\mu} & = & \frac{2 \pi}{L_{\mu}} \, L_{\nu} \,
(1-x_{\nu}) \,,\quad
\tilde{\theta}_{{\vec x},\rho} = 0 \,,\,\,\, \rho \neq \mu, \nu \, .
\label{def:global_step}
\eeqn

The acceptance rate of the global step changes with $\beta$ in a different
way, depending on the lattice geometry.
In our $T=0$ studies (on $32^3$ lattices) we found
that the acceptance of global update steps
drops (more rapidly than
exponentially) from 0.48 at $\beta=1.0$ to 0.0056 at $\beta=2.0$.
For higher $\beta$ essentially
no global offers are successful.

In our $T \ne 0$ studies on $32^2 \times 8$ lattices, however, we found
the acceptance changing smoothly (nearly exponentially, even across the
deconfining transition) from 0.58 at $\beta=1.0$ to 0.18 at $\beta=3.0$.
A closer look reveals that the higher acceptance rate is due to more
frequent global changes of the flux penetrating the 12 plane (i.e.
magnetic flux direction).

In summary, one total Monte Carlo update cycle consists of three cycles
of local update, each consisting of a Metropolis sweep followed by a
microcanonical sweep, interchanging with a global update as described above.

In the finite temperature case, the measurement of $D_T$ turns out to be
highly sensitive with respect to the insufficient removal (by the gauge
fixing procedure and its repetitions, see the next subsection) of Dirac
strings wrapping around the third direction. This is a case where the results
with and without global updates, mainly adding and subtracting fluxes through
the $12$ plane, differ. We comment on this problem and how to deal
with it in Section~\ref{sec:finite_temperature}.

\subsection{Landau Gauge Fixing}
\label{subsec:LandauGauge}

The Landau gauge was chosen first of all because it is
the most popular gauge to define a gauge field propagator.
In this gauge the gauge propagator (\ref{def:propagator})
is expected to satisfy
the transversality condition
\beqn
F(q^2) = q_{\mu}~D_{\mu\nu}({\vec q})~q_{\nu} \equiv 0 \; .
\label{def:transversality}
\eeqn
In the case of zero temperature, for example, this allows to describe
the propagator by a single function $D(q^2)$ alone, defined by
\beqn
D_{\mu\nu}({\vec q})=\Bigl( \delta_{\mu\nu} - \frac{q_\mu~q_\nu}{q^2}
\Bigr)~D(q^2) \; .
\label{def:projector}
\eeqn
We will see that for any practical implementation of the Landau gauge
(\ref{def:transversality}) is slightly violated. This degree of violation
can, however, be easily controlled by sharpening the convergence criteria
of the gauge--fixing algorithm.
More important is the remark that in the case of the {\it angle}--definition
of the vector potential (\ref{def:angle}) it is really necessary to select the
transverse part by projection using (\ref{def:F_project_2}) and
(\ref{def:D_project_2}). Then it is interesting to see
where ({\it e.g.} in momentum space) the violation of transversality,
quantified by the longitudinal propagator $F$, is coming from.

There is a second reason to choose the Landau gauge.
We intend to split the gauge field into a regular (photon)
and a singular (monopole) part by reconstructing the field due to the Dirac
plaquettes which, on the other hand, are forming the monopoles.
This reconstruction becomes unique in the Landau gauge.

In order to implement the gauge fixing condition
(\ref{def:cos_gauge_condition}) we have chosen
a mixture of overrelaxation and non-periodic
gauge transformations~\cite{Bogolubsky:1999},
both applied in alternating order.

Iterative overrelaxation has to be practiced in a checkerboard
fashion. Starting, say, with the odd sublattice we have first to
find for each odd site ${\vec n}$ a suitable $\omega_{\vec n}$ which
maximizes the following function of $\omega$
\beqn
{\cal G}^{loc}_{{\vec n}}(\theta,\omega) = \sum_{\mu}
\left( \cos (\theta_{{\vec n},\mu} - \omega_{{\vec n}})
     + \cos (\theta_{{\vec n}-{\vec \mu},\mu} + \omega_{{\vec n}})
     \right) \, ,
\label{def:local_gauge_condition}
\eeqn
which represents the part of ${\cal G}$ actually depending on
$\omega_{\vec n}$,
and second to perform immediately the updatings of the neighboring
link angles
\beqn
\theta_{{\vec n},\mu} & \rightarrow &
\theta_{{\vec n},\mu} - \omega_{{\vec n}} \nonumber \\
\theta_{{\vec n}-{\vec \mu},\mu} & \rightarrow &
\theta_{{\vec n}-{\vec \mu},\mu} + \omega_{{\vec n}}\,.
\label{def:gauge_trafo}
\eeqn
This can be done simultaneously for half of the sites,
namely ${\vec n} \in \Lambda_{odd}$.
Afterward the same procedure is applied to the even sublattice.
One odd/even pair of gauge updates constitutes
one single iterative overrelaxation
step. Each overrelaxation iteration is followed by a zero mode subtraction
(to be explained in the next subsection).

The angle $\omega_{\vec n}$ can be easily found as
\beqn
\tan(\omega_{\vec n}) =  \frac
{\sum_{\mu} \Bigl( \sin(\theta_{{\vec n},\mu})
- \sin(\theta_{{\vec n}-{\vec \mu},\mu})\Bigr) }
{\sum_{\mu} \Bigl( \cos(\theta_{{\vec n},\mu})
+ \cos(\theta_{{\vec n}-{\vec \mu},\mu})\Bigr) }\; .
\label{def:right_omega}
\eeqn
These gauge angles are multiplied by the overrelaxation factor,
$\omega_{\vec n} \to \eta \omega_{\vec n}$ and bounded
by $|\omega_{\vec n}| < \pi$ before the iteration
(\ref{def:gauge_trafo}) is performed on all links.
A good overrelaxation parameter was found to be $\eta=1.8$.
According to our experience from studies on $16^3$ lattices
this $\eta$ leads to fastest convergence, almost independently of $\beta$.
This value has then been applied for all iterative gauge fixings.

The overrelaxation will
usually  be stopped if in the last overrelaxation
step the average increase of the gauge functional ${\cal F}$
[Eq.(\ref{def:cos_gauge_functional})] per link
is found to be less
than $10^{-6}$. After this was discovered,
the gauge fixing procedure ends with a final
zero mode subtraction (see below).

At any {\it local extremum} of ${\cal F}$ [Eq.(\ref{def:cos_gauge_functional})]
the following condition would be satisfied
everywhere on the even {\it and} odd
sublattice:
\beqn
\left( \partial_{\mu} A_{\mu} \right)_{\vec n} \equiv
\sum_{\mu} \Bigl( A_{{\vec n}+\frac{1}{2}{\vec \mu},\mu}
                - A_{{\vec n}-\frac{1}{2}{\vec \mu},\mu} \Bigr) \equiv
\frac{1}{g~a}~ \sum_{\mu} \Bigl(\sin(\theta_{{\vec n},\mu})
                    - \sin(\theta_{{\vec n}-\hat{\mu},\mu})\Bigr) = 0 \; .
\label{def:A_divergence}
\eeqn
Having the vector potential localized on the midpoints of links,
its divergence is naturally defined on sites ${\vec n}$.
Exact vanishing of the divergence of $A_{\mu}$ can be expected in the result
of Landau gauge fixing only for the {\it sine}--definition
of the vector potential $A_{\mu}$.

The algorithm outlined above will
not in general lead to the
absolute (global) maximum of the gauge functional ${\cal F}$
(\ref{def:cos_gauge_functional}). Typically it will get stuck in
one of the local maxima, which are called Gribov copies of the
true maximum. This is the so-called Gribov problem. It is
partially cured by repeating the same gauge fixing procedure,
applying it to random gauge copies of the original Monte Carlo
configuration $\theta^{MC}$, assuming that one of these copies
might be situated in the basin of attraction of the true maximum.
The number of gauge equivalent configurations produced to restart
the gauge fixing is denoted as $N_G$, and the iterative gauge
fixing generically leads to really different maxima. We have then
to content with the {\it best out of all $N_G+1$ local maxima} of
the gauge functional. The convergence, with increasing $N_G$, of a
gauge dependent quantity evaluated on a given gauge field ensemble
with the help of the best Gribov copy gives an indication
of the sensitivity of this quantity with respect to the
misidentification of the true maximum. We have applied this
philosophy to two sets of data, the propagator data themselves at
large or small momenta, and to the fit parameter emerging from a
fit of the gauge boson propagator. One should not be surprised
that the number $N_G$, which is necessary to achieve uniform
convergence of the propagator in momentum space and/or of the fit
parameters, differs strongly between zero and finite temperature.
At finite temperature there are strong differences between $D_L$
and $D_T$.

In section \ref{sec:analytical_consideration} we have stressed the importance
to find a local maximum of (\ref{def:cos_gauge_functional}) accompanied by a
minimal length of Dirac strings. Within our implementation, the start from
a new random gauge copy is done in the hope in producing a new Gribov copy
reachable from the previous ones only by a discrete gauge transformation.
We have monitored the number of Dirac strings in each of the local
maxima of the gauge functional ${\cal F}$ (\ref{def:cos_gauge_functional}).
Each time the recent best value of the gauge functional was replaced by a
better (higher) one, the number $N_D$ of Dirac strings detected in the
correspondingly best Gribov copy decreased compared to the previous one.

\begin{figure}[!htb]
\begin{center}
\begin{tabular}{cc}
\epsfxsize=6.0cm \epsffile{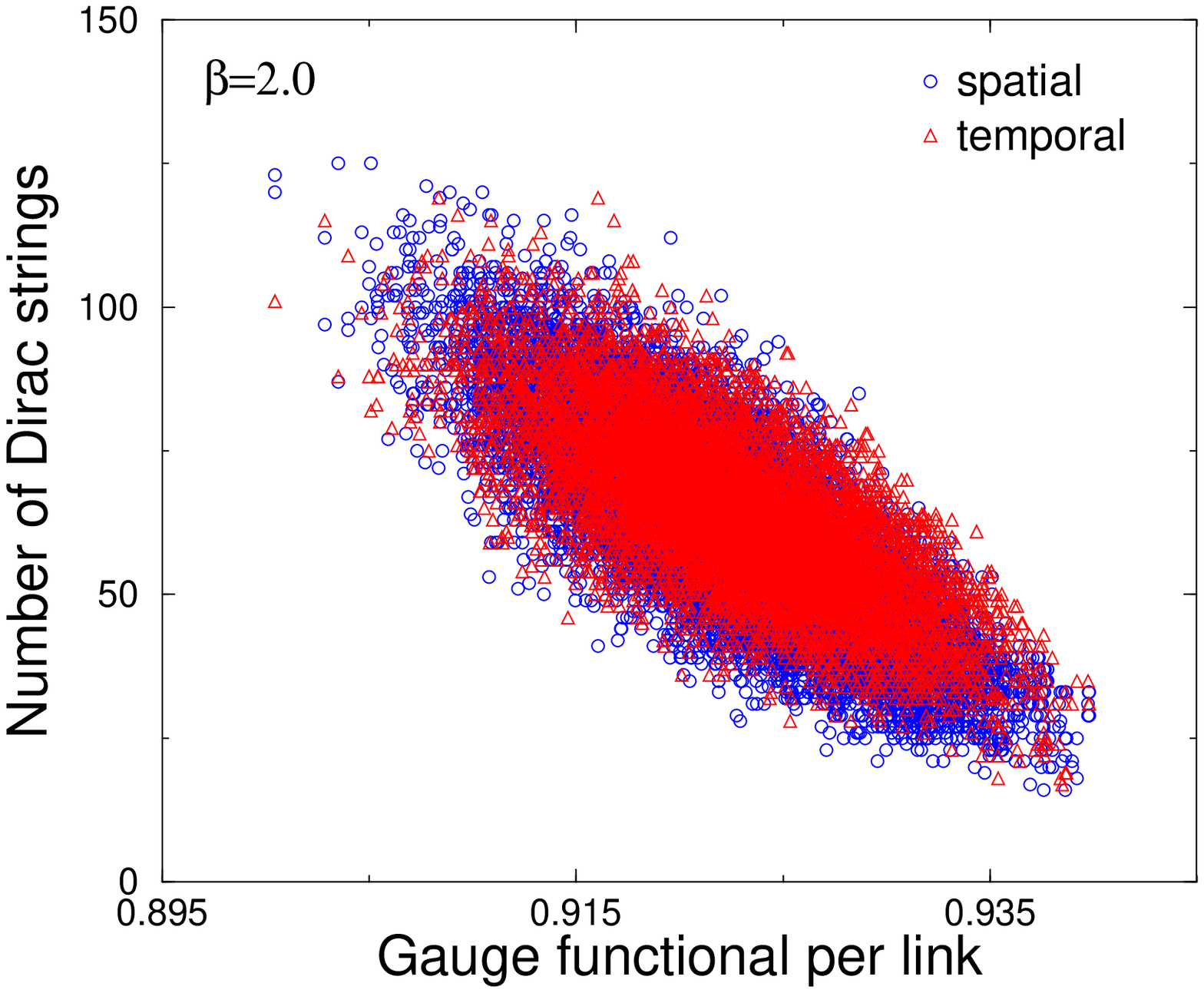} & \hspace{5mm}
\epsfxsize=6.0cm \epsffile{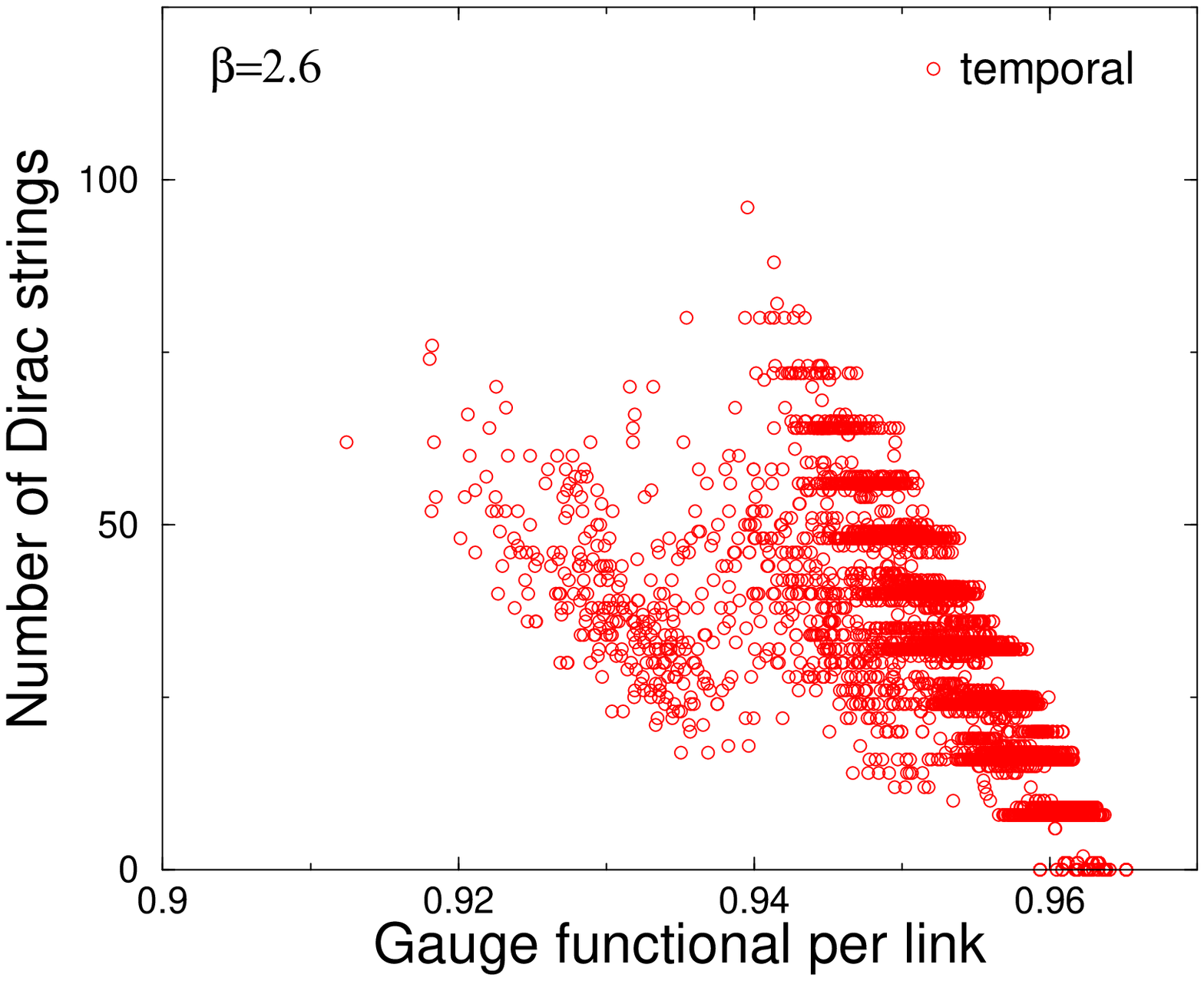}  \\
(a) & \hspace{5mm} (b) \\ \\
\epsfxsize=6.0cm \epsfysize=4.85cm \epsffile{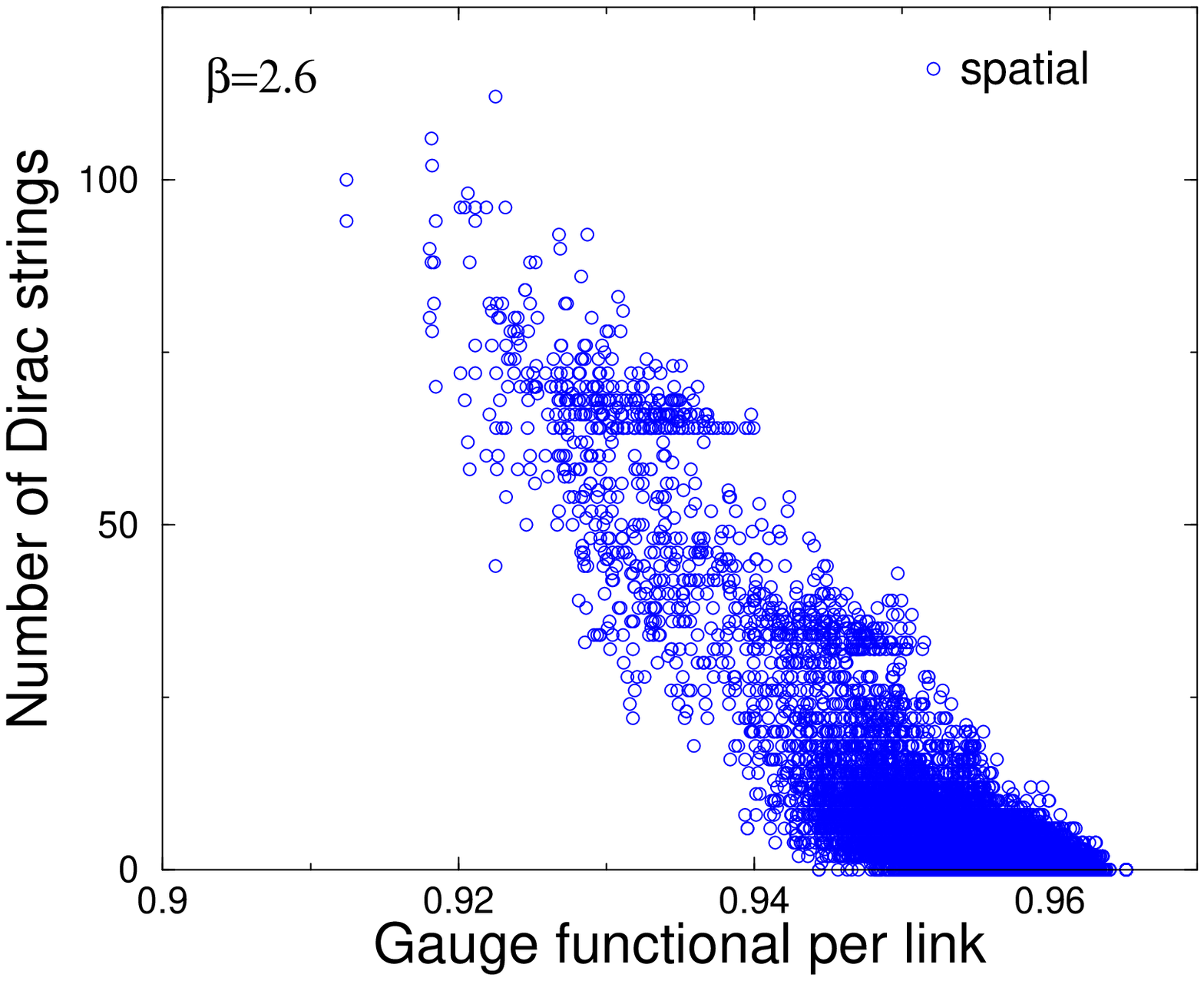}  &
\hspace{5mm}
\epsfxsize=6.0cm \epsffile{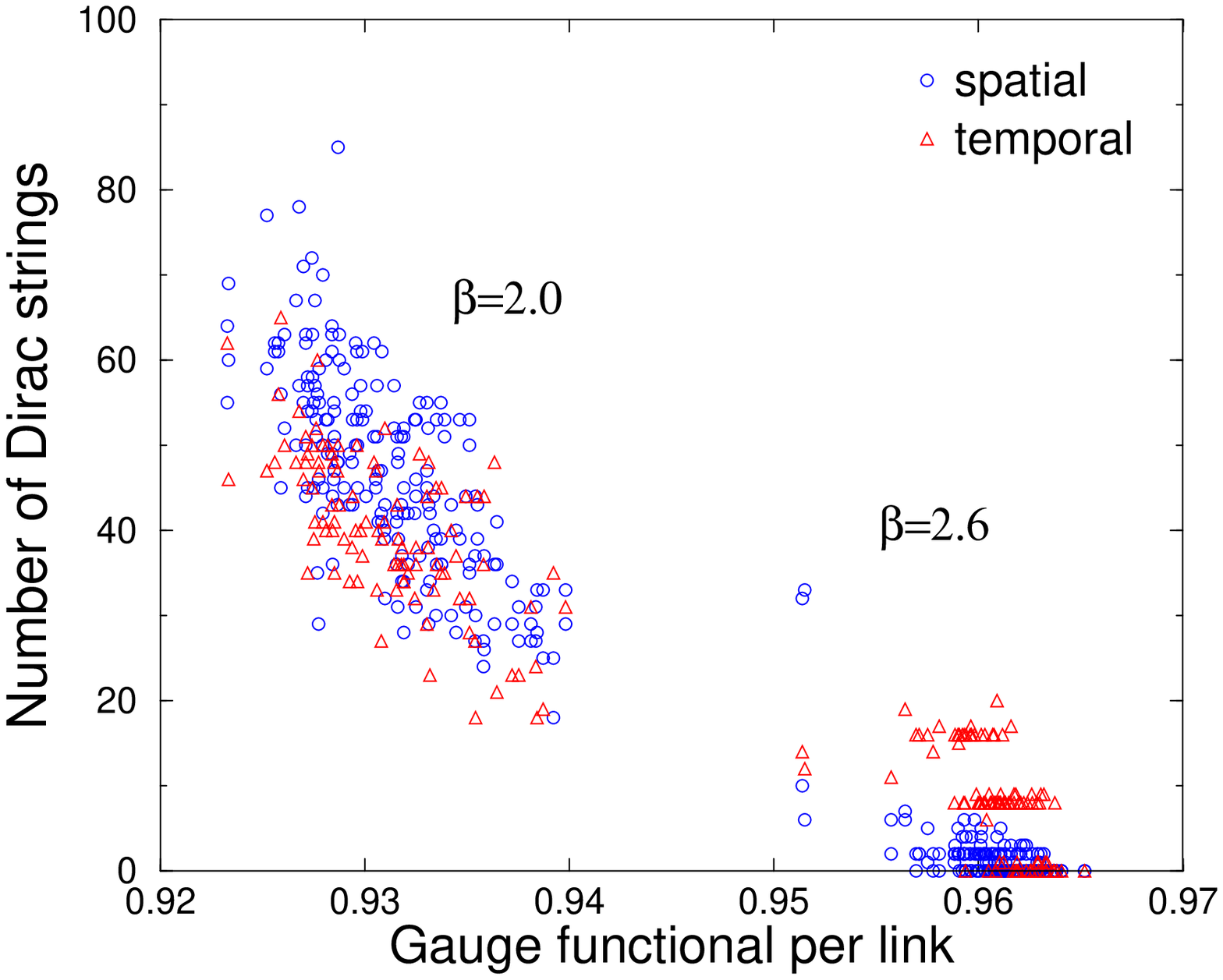} \\
(c) & \hspace{5mm} (d)
\end{tabular}
\end{center}
\vspace{-5mm}
\caption{Scatter plots of the average gauge functional ${\cal F}(\theta)$
         per link and the related number
         of Dirac strings in the
         {\it temporal} ($N^D_3$) and/or the {\it two spatial}
         ($N^D_1$ and $N^D_2$) directions for 100
         configurations on a lattice $32^2 \times 8$:
         (a) $N^D_3$, $N^D_1$ and $N^D_2$ vs ${\cal F}(\theta)$
             in the confinement phase ($\beta=2.0$), each configuration
             represented by 101 Gribov copies;
         (b) $N^D_3$ vs ${\cal F}(\theta)$ in the deconfinement phase
             ($\beta=2.6$);
         (c) same as in (b) for $N^D_1$ and $N^D_2$ ;
         (d) the two samples represented only by the best out of 101 Gribov
           copies.}
  \label{fig:F_ND_scatter}
\end{figure}

In Fig.~\ref{fig:F_ND_scatter}
we show scatter plots relating the gauge
functional ${\cal F}$ to the number of Dirac strings per direction $N^D_i$
($i=1,..,3)$) for two thinned ensembles of 100 out of 1000 equilibrium
configurations  on $32^2 \times 8$ lattices, each of them entering with 101
Gribov copies (local maxima of ${\cal F}$).
Fig.~\ref{fig:F_ND_scatter}(a) refers to $\beta=2.0$ (confined phase) and
shows all local maxima (Gribov copies). The isotropy of Dirac strings
is clearly visible. Fig.~\ref{fig:F_ND_scatter}(b) shows how in the
deconfined phase the number $N^D_3$ of temporal Dirac strings is correlated
with the locally maximal value of ${\cal F}$.
Fig.~\ref{fig:F_ND_scatter}(c) shows it for the $N^D_1$ and $N^D_2$
(number of spatial Dirac strings). In both cases a slight tendency to
clustering near multiples of 8 or 32 (the respective periodicity
of the lattice) is
visible, in particular for the highest values of ${\cal F}$.
The selection of the best copies restricts the ensemble to bigger ${\cal F}$
and smaller $N^D_i$ as shown in Fig.~\ref{fig:F_ND_scatter}(d).
In the confinement phase the Dirac strings are
 still approximately
isotropic. In the deconfined phase, however, the highest
 values of ${\cal F}$
are correlated with low multiplicity of $N^D_1$ and $N^D_2$,
and $N^D_3$ corresponding to zero, with
one or a few periodically winding Dirac strings
more frequently in the temporal direction.

We stress, that the configurations of which the gauge fixing was
investigated here in some detail, had been produced with the update algorithm
including global updates. Winding Dirac strings of some life time are also
produced in ensembles that
are generated {\it without} global updates, but
less frequently. Therefore, the inefficiency of the random gauge transformation
in exploring
more of the gauge orbit is a handicap also if global updates
are suppressed. We will see later that certain problems which show up
in the finite temperature propagator $D_T$ can be ameliorated, but not
completely cured, by abandoning global updates and increasing the number
of Gribov copies.

We had not the opportunity to reconstruct exactly the spatial conformation
of the Dirac strings in each gauge copy.
But for each gauge copy we can determine whether
multiple
Dirac strings (running along a certain direction
with same or different orientation)
can be definitely excluded or not.  At low $\beta$ this can never be
excluded, but at high $\beta$ this could
be used as an additional criterion to reject Gribov copies which
contain double Dirac strings
(in case of two {\sl oppositely} oriented strings).

Although the stopping criterion was formulated for the {\it global
increase} of ${\cal F}$ we did not find a systematic {\it local variation},
depending on the distance from a monopole, of the violation of
(\ref{def:A_divergence}), expressed by the quantity
\beqn
\left( \partial_{\mu} A_{\mu} \right)^2_{\vec n} \equiv
\left\{ \sum_{\mu} \Bigl( A_{{\vec n}+\frac{1}{2}{\vec \mu},\mu}
         - A_{{\vec n}-\frac{1}{2}{\vec \mu},\mu} \Bigr) \right\}^2 \, .
\label{def:A_divergence_violation}
\eeqn
This suggests that the differential gauge condition (\ref{def:A_divergence})
is uniformly approaching zero.

\begin{figure}[!htb]
  \begin{center}
  \epsfxsize=8.0cm \epsffile{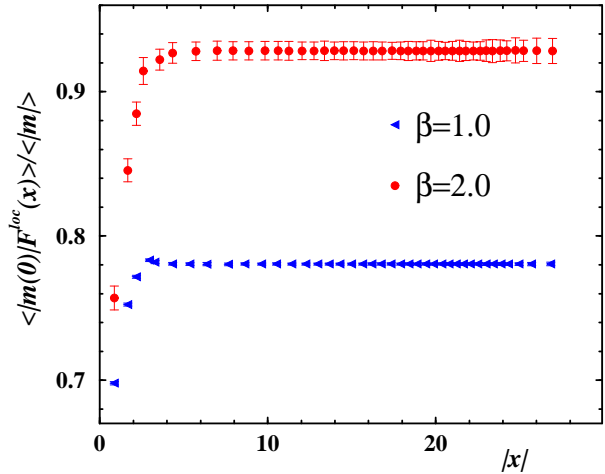}
  \end{center}
  \vspace{-5mm}
  \caption{Suppression of the local gauge functional
  ${\cal F}^{loc}_{{\vec n}}$
  near to a monopole for $\beta=1.0$ and $\beta=2.0$.
  For better presentation part of the data points are not plotted.
  Notice the $\beta$ dependence of the bulk average
  of ${\cal F}^{loc}_{{\vec n}}$.}
  \label{fig:F_loc_suppression}
\end{figure}

In contrast to this, we found a systematic local variation of
the local gauge functional itself,
\beqn
{\cal F}^{loc}_{\vec n}(\theta) =
\sum_{\mu} \left(
                  \cos ( \theta_{{\vec n},\mu} )
                + \cos ( \theta_{{\vec n}-{\vec \mu},\mu} ) \right) \, ,
\label{def:local_gauge_functional}
\eeqn
with the distance from
a monopole. Near to a monopole, its value is suppressed compared to
the bulk average. This is illustrated in Fig.~\ref{fig:F_loc_suppression}
for $\beta=1.0$ and $\beta=2.0$.

\subsection{Zero Mode Subtraction}
\label{subsec:remov_zero_modes}

There are certain modes of the gauge field which are not suppressed
by the action. If they are not taken care of properly, the gauge field
propagator is known to be spoiled~\cite{Bogolubsky:1999}. For instance,
adding some constant to all link angles $\theta_{{\vec n},\mu}$ in one
particular direction does not change the action (\ref{def:action}), as
neither does adding a multiple of $2\pi$ to one particular link angle.
Non-periodic gauge transformations can be considered to implement these
changes. We have used it in our simulations to immediately eliminate zero
modes related to the appearance of the volume average of the link angles
in a given direction
\beqn
\overline{\theta_{\mu}}=\sum_{{\vec n}} \theta_{{\vec n},\mu}/|\Lambda| \, ,
\label{def:zero_modes}
\eeqn
where $|\Lambda|$ is the lattice volume.
We subtract this volume average from each link angle $\theta_{{\vec n},\mu}$
in $\mu$ direction,
\beqn
\theta_{{\vec n},\mu} \to \theta_{{\vec n},\mu} - \overline{\theta_{\mu}}
+ 2\pi k \, ,
\label{def:zero_mode_subtraction}
\eeqn
with integer $k$ chosen such that $\theta^{g}_{{\vec n},\mu} \in (-\pi,\pi]$.
This operation can be imagined as resulting from a gauge function
$g_{\vec n} = -~\overline{\theta_{\mu}} n_{\mu}$ (no summation!)
which is in general non-periodic.
This special sort of gauge fixing is completed when
zero modes in all $d$ directions were subtracted.

The zero-mode subtraction step is applied after each
overrelaxation step. Thus it is always performed before the
measurements are done on the gauge-fixed configuration.

In our runs we measured the photon propagator after each 10th total Monte
Carlo update cycle to avoid
autocorrelations as much as possible.
Typically we used  $500$ gauge-fixed configurations per data point for the
$32^3$ lattice  and about $2000$ configurations for the $32^2\times8$ lattice.


\section{The zero-temperature propagator in Landau gauge}
\label{sec:zero_temperature}

In coordinate space, the gauge boson propagator that is studied, is read
\beqn
D_{\mu\nu}\left({\vec m}+\frac{1}{2}{\vec \mu}-{\vec n}-\frac{1}{2}{\vec \nu} \right)
= \langle A_{{\vec m}+\frac{1}{2}{\vec \mu},\mu}
          A_{{\vec n}+\frac{1}{2}{\vec \nu},\nu} \rangle
\label{def:prop_in_coord_space}
\eeqn
with the {\it angle}-- or {\it sine}--definition of $A$.
For brevity, we will refer to this propagator later in momentum space
as $D^{ang}_{\mu\nu}$ or $D^{sin}_{\mu\nu}$.
The two scalar functions occurring in momentum space
(\ref{def:tensor_structure})
we will denote as $D^{ang}$ and $F^{ang}$ or $D^{sin}$ and $F^{sin}$,
respectively.
The last one, $F^{sin}$ should vanish in the Landau gauge.
We have observed
that this is indeed the case with an accuracy determined by
the stopping criterion of the iterative overrelaxation.
For the {\it sine}--propagator the transverse part could be calculated
directly
just by evaluating and summing the diagonal components appearing in
(\ref{def:D_project_2}). When the longitudinal propagator vanishes only {\it
approximately}, one can extract the transverse propagator following
(\ref{def:F_project_2}, \ref{def:D_project_2}).

\begin{figure}[!htb]
  \begin{center}
    \begin{tabular}{cc}
       \epsfxsize=6.0cm \epsffile{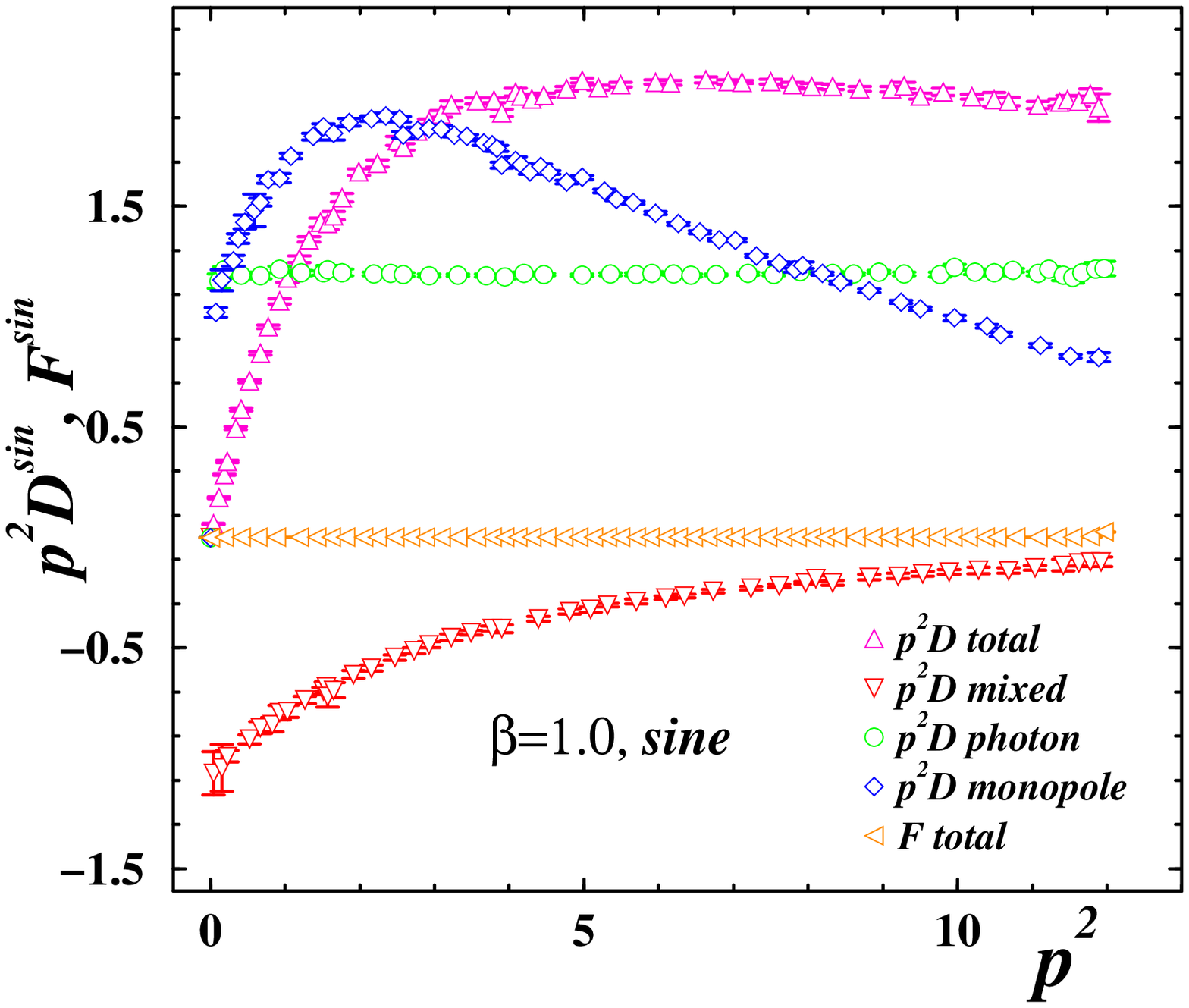} &\hspace{5mm}
       \epsfxsize=6.0cm \epsffile{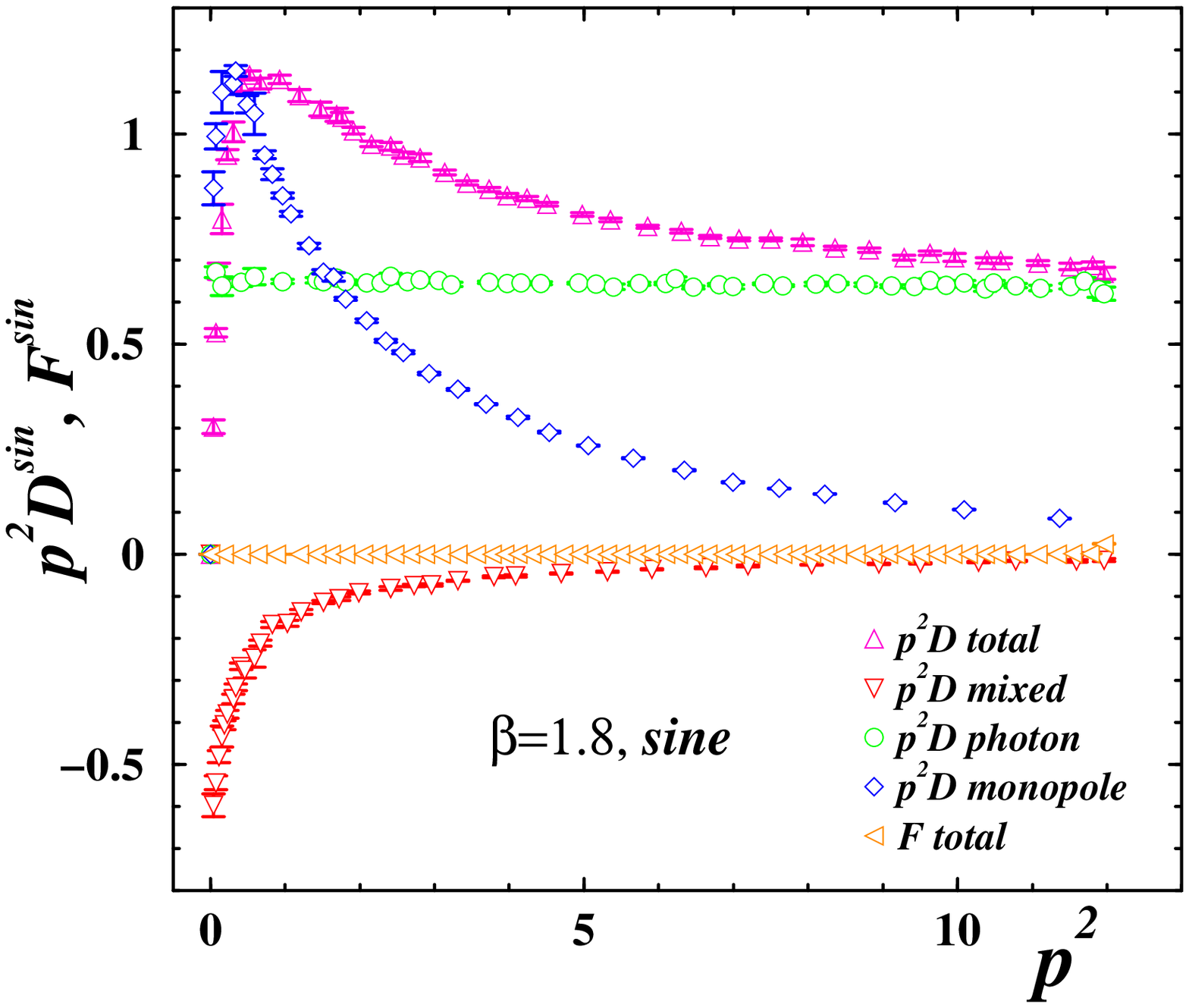} \\
        (a) & \hspace{5mm}(b)
     \end{tabular}
  \end{center}
  \vspace{-5mm}
  \caption{The Landau gauge {\it sine}--propagator
      as a function of $p^2$ measured on a $32^3$ lattice :
    for the transverse part, $p^2 D^{sin}$
    ((a) at $\beta=1.0$ and (b) at $\beta=1.8$)
    we show the full propagator, the singular (mono), the regular (phot)
    and the mixed contribution for comparison. In addition, we show the
    (vanishing) longitudinal propagator $F^{sin}$.
    The data represent the evaluation of $N_G=20$ gauge copies.
  For better presentation part of the data points are not plotted.}
  \label{fig:momdep_sin_propagator}
\end{figure}

\begin{figure}[!htb]
  \begin{center}
     \begin{tabular}{cc}
        \epsfxsize=6.0cm \epsffile{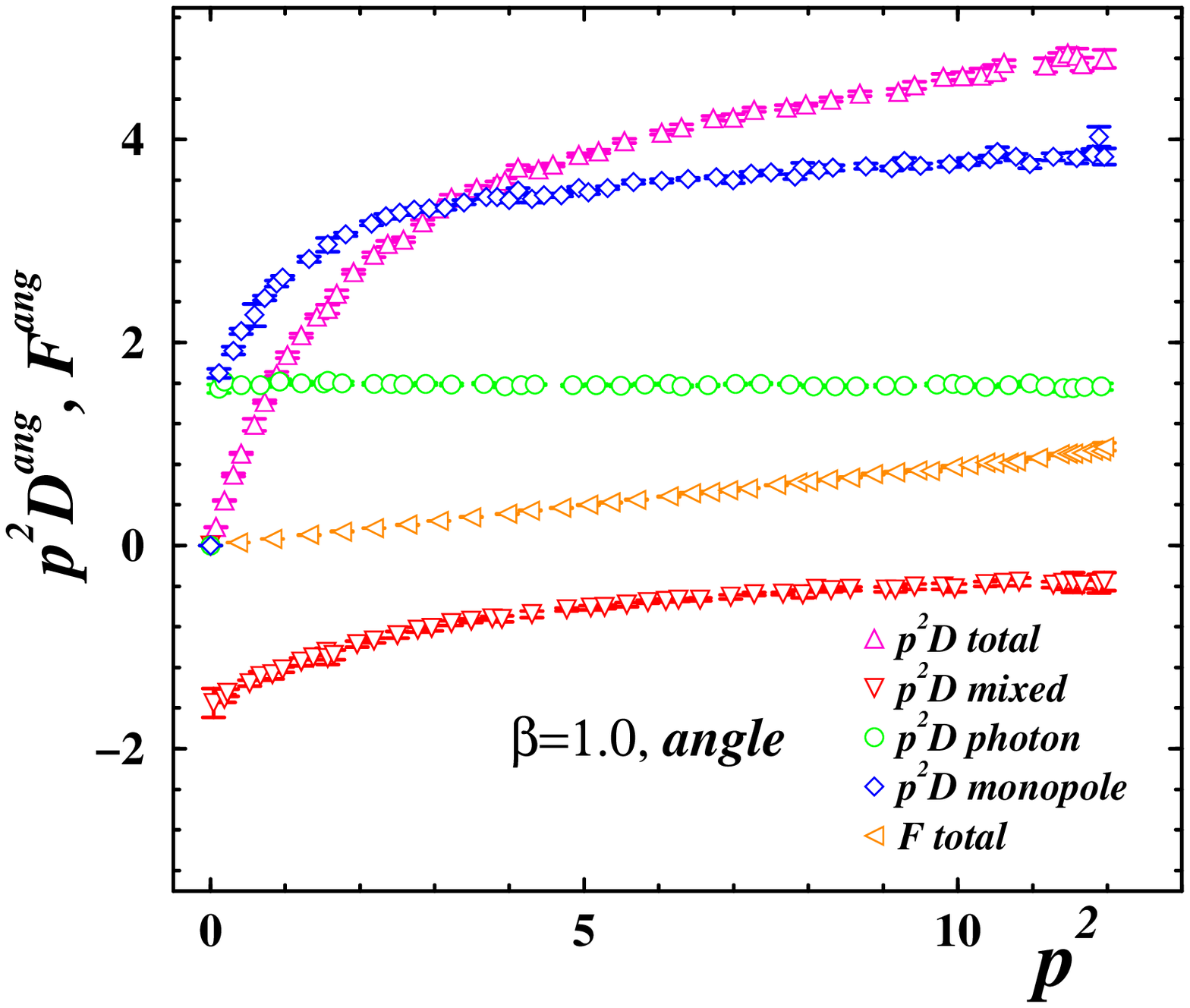}  & \hspace{5mm}
        \epsfxsize=6.0cm \epsffile{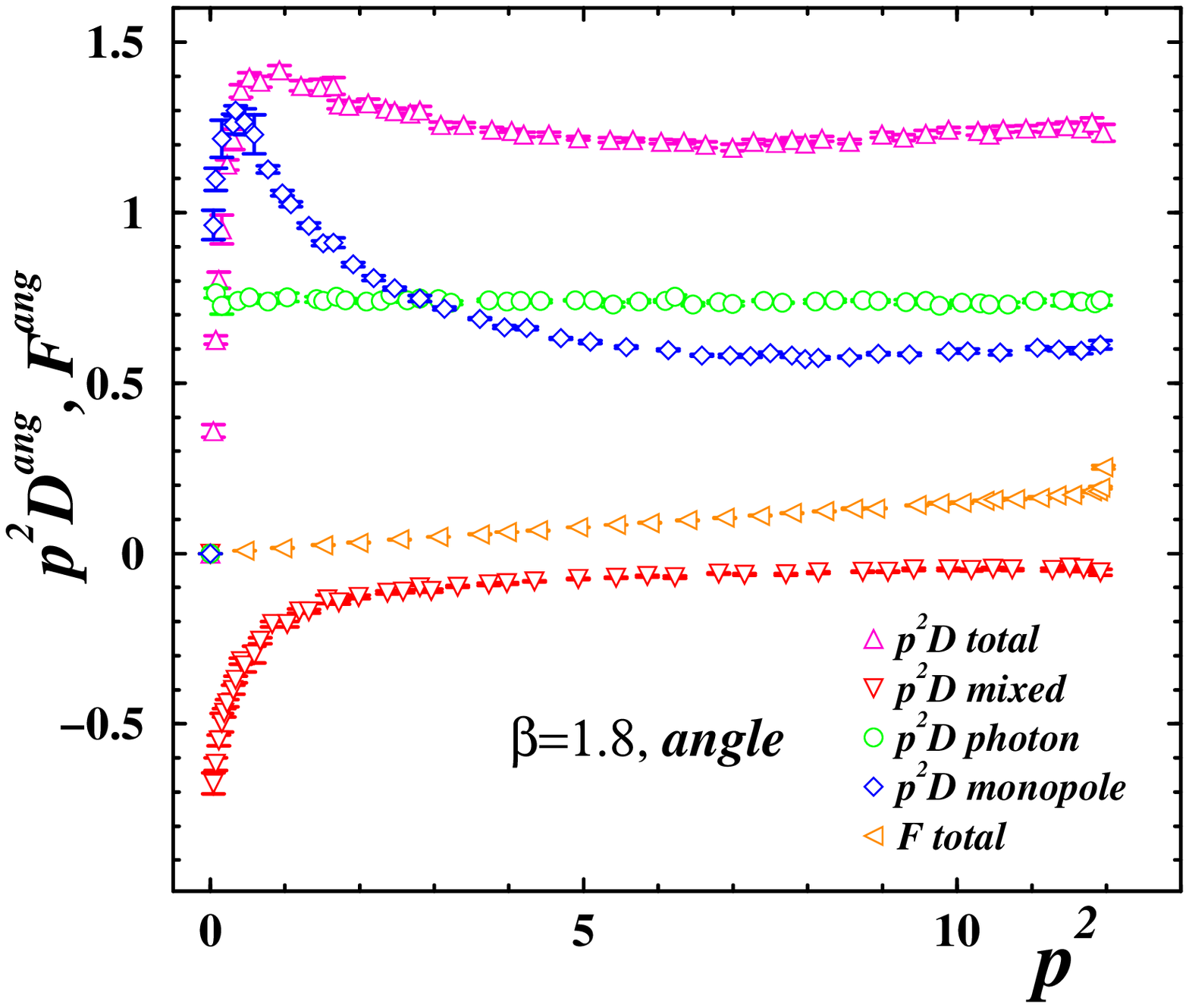}  \\
        (a) & \hspace{5mm} (b)
     \end{tabular}
  \end{center}
  \vspace{-5mm}
  \caption{The same as in Fig.~\ref{fig:momdep_sin_propagator}
           for the Landau gauge {\it angle} propagator. Notice
           the nonvanishing longitudinal propagator $F^{ang}$
           decreasing with growing $\beta$.}
  \label{fig:momdep_ang_propagator}
\end{figure}

In Fig.~\ref{fig:momdep_sin_propagator}
we show the different forms of the transverse propagator $D^{sin}$ and its
components as well as the vanishing longitudinal propagator $F^{sin}$.
These data were obtained on a $32^3$ lattice for (a) $\beta=1.0$ and (b)
$\beta=1.8$, with $N_G=20$ Gribov copies evaluated in addition to each
original Monte Carlo configuration. The data at other $\beta$ were
produced under the same conditions.
In Fig.~\ref{fig:momdep_ang_propagator}
the same is presented
for $D^{ang}$, its components and non--vanishing $F^{ang}$.

For the {\it angle}--definition of $A_{\mu}$, the decomposition
into components is strictly additive
(\ref{def:D_F_decomposition}). We have observed that the
longitudinal propagator $F^{ang}$ and its components are well
described by a form $F(p^2) = P~p^2$, where $P$ is a constant. We
find that $F^{ang,phot}$
essentially
coincides with the full $F^{ang}$. The
size of $F^{ang}$ and its photon component is of the same order of
magnitude as the transverse part at $\beta=1.0$, while the
monopole and mixed part are one order of magnitude smaller. At
$\beta=1.8$, the size of $F^{ang}$ and its photon component is an
order of magnitude smaller than at $\beta=1.0$, while the monopole
and mixed part are negligible.

For the {\it sine}--definition of $A_{\mu}$, the monopole part of the
transverse propagator $D^{sin,mono}$ has a maximum in the low momentum
region (which moves more and more towards $p^2=0$ with higher $\beta$)
before it drops towards $p^2=0$. For the {\it angle}--definition,
a maximum of $D^{ang,mono}$ develops only for $\beta > 1.0$.

Summarizing, we have observed that the
longitudinal part $F$ is either
zero (for the $\sin\theta$ definition of the gauge field in the
correlator) or it is non-zero, and then it coincides with its
photon part (for the $\theta$ definition).
Therefore,
at zero temperature the Landau gauge propagator
$D^{ang}_{\mu\nu}$ is not completely transverse. This is entirely
due to the difference between the definitions of the vector potential.
This discrepancy, expressed by the nonvanishing $F^{ang}$, becomes
ameliorated at higher $\beta$.

For both definitions of $A_{\mu}$ we find that the regular (photon)
part of the transverse propagator is singular at $p^2 \to 0$
like $D^{phot} \sim 1/p^2$,
while the full transverse propagator is not.
Following Ref.~\cite{CISLetter} we try to describe the two by functions
of the form
\beqn
D(p^2) = \frac{Z}{\beta}\frac{m^{2\alpha}}{p^{2(1+\alpha)}+m^{2(1+\alpha)}}
+ C \,
\label{def:anomalous_fit}
\eeqn
and
\beqn
D^{phot}(p^2) = \frac{Z^{phot}}{\beta}\frac{1}{p^2} + C^{phot} \, .
\label{def:regular_fit}
\eeqn
The model function (\ref{def:anomalous_fit}) is similar to some of the
functions discussed in Refs.~\cite{CurrentQCD,Ma} where the propagator
in gluodynamics was studied.
In the case of $T=0$ we expect these two curves to differ {\it at all $\beta$}
in order to accommodate the (permanent) confinement property of the model.

\begin{figure}[!htb]
  \begin{center}
  \begin{tabular}{cc}
  \epsfxsize=6.0cm \epsffile{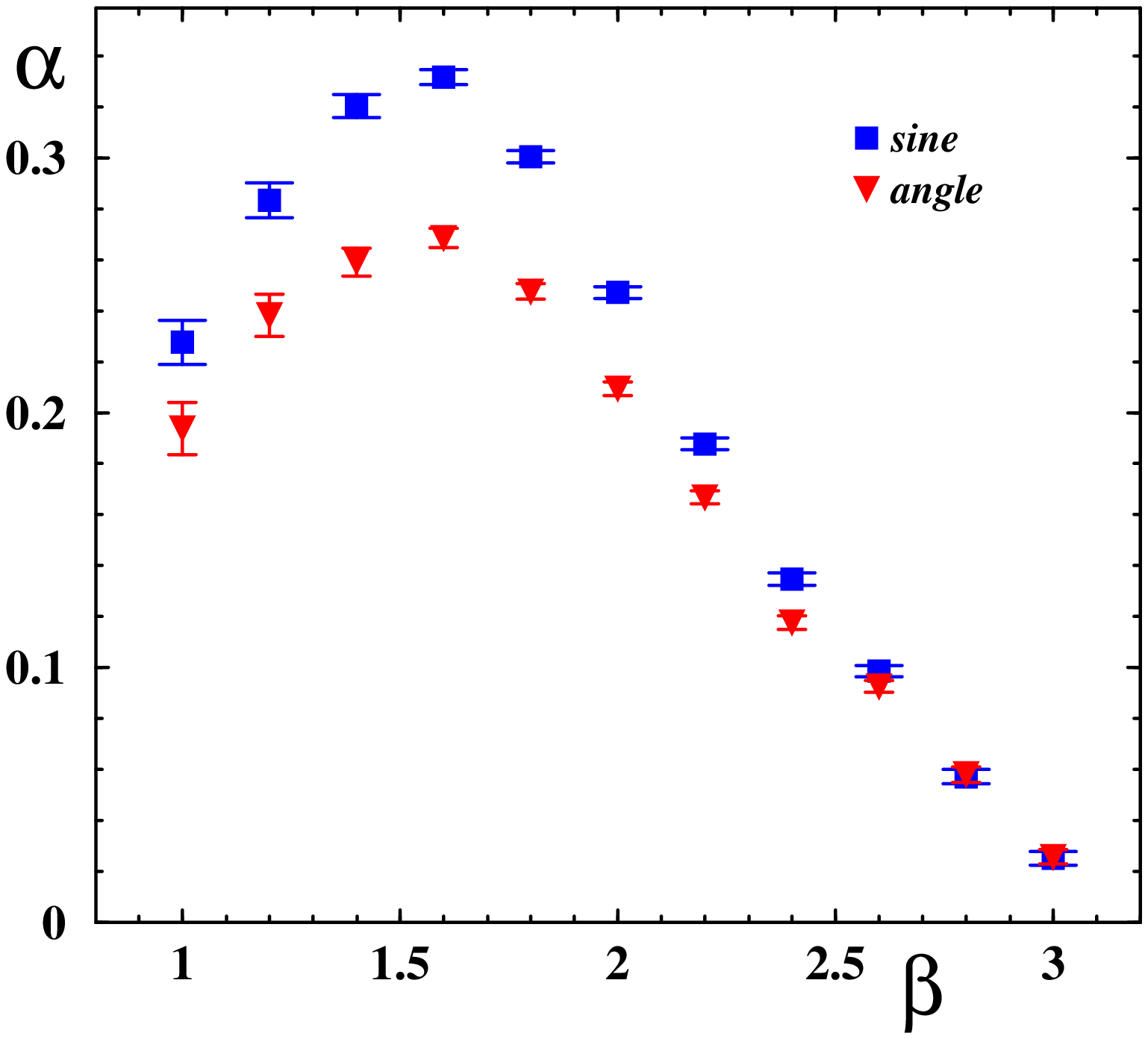}  & \hspace{5mm}
  \epsfxsize=6.0cm \epsffile{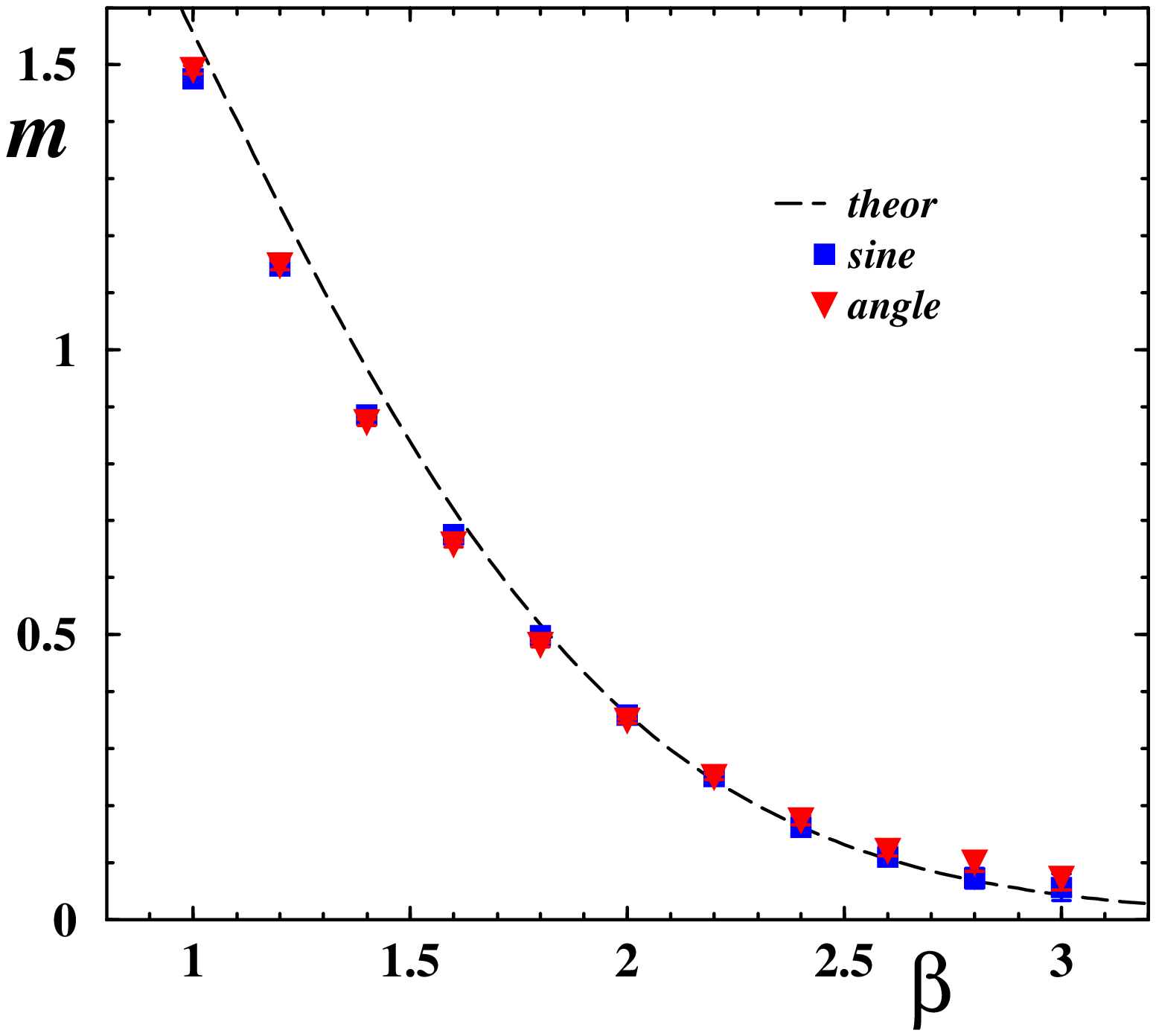} \\
  (a) & \hspace{5mm} (b)\\
  \epsfxsize=5.7cm \epsffile{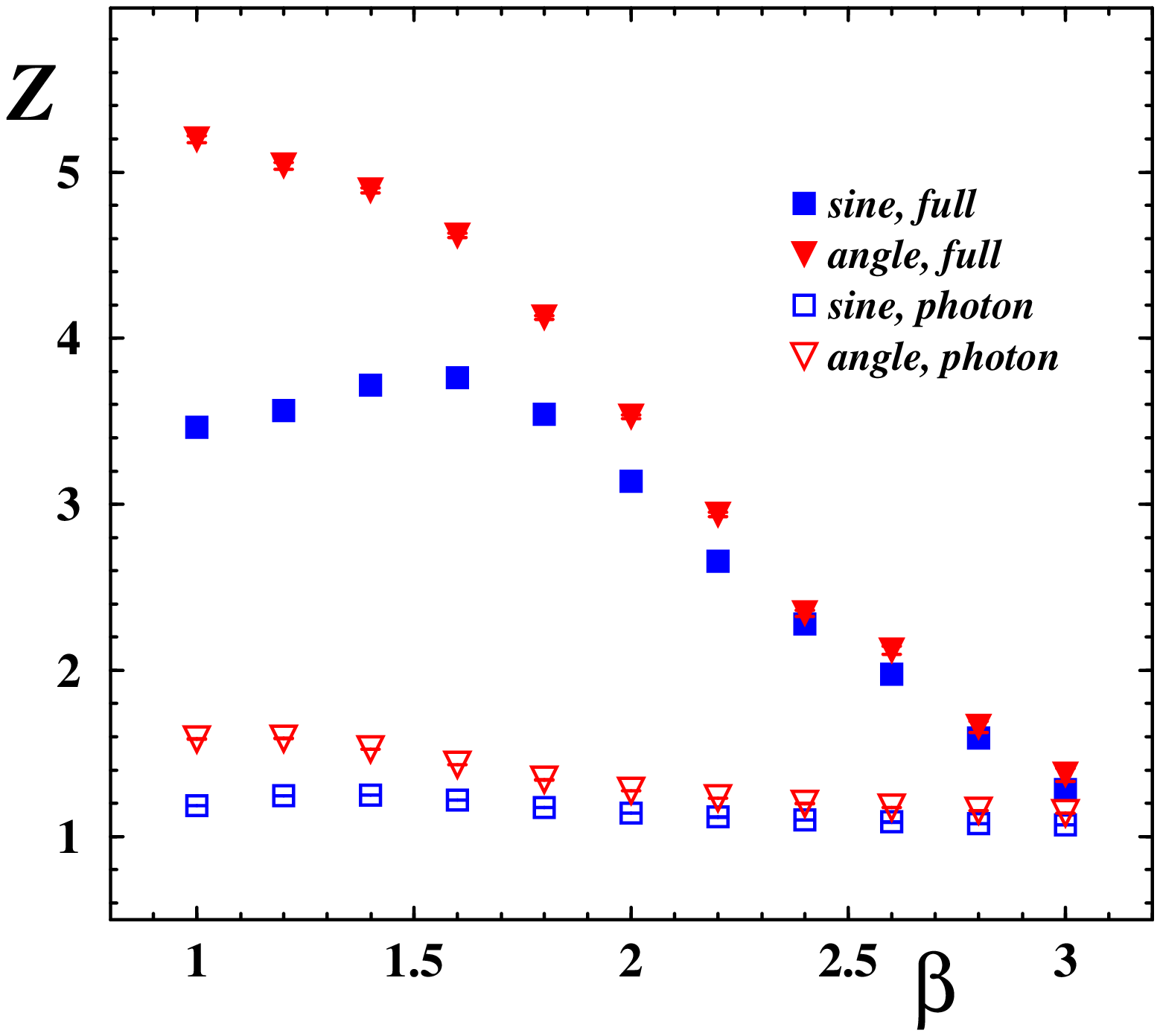}  & \hspace{5mm}
  \epsfxsize=6.0cm \epsffile{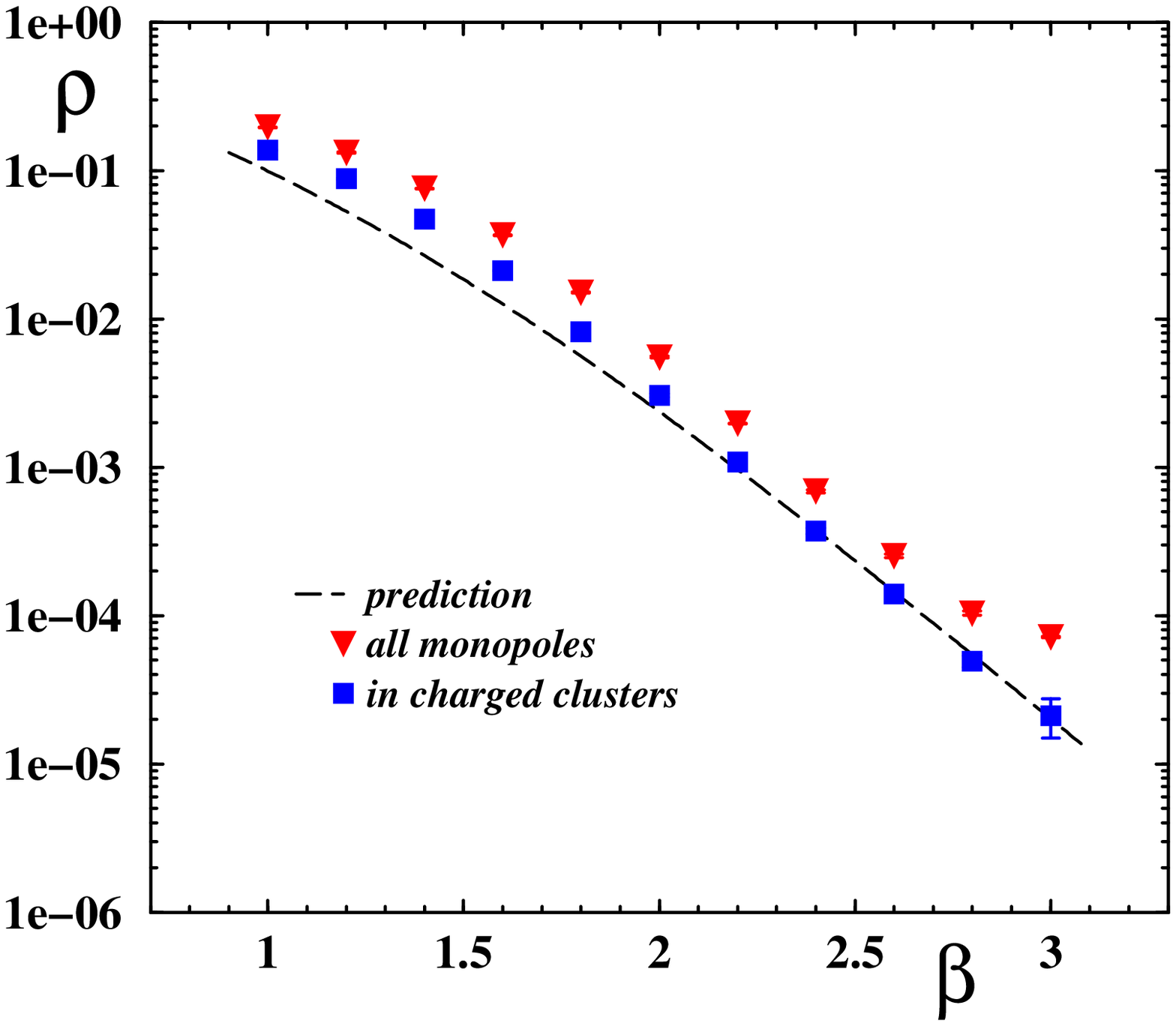} \\
  (c) & \hspace{5mm} (d) \\
  \end{tabular}
  \end{center}
  \vspace{-5mm}
  \caption{The best fit parameters for the
     zero temperature {\it sine}-- and {\it angle}--propagators
     as functions of $\beta$:
     (a) the anomalous dimension $\alpha$,
     (b) the mass parameter $m$ with the theoretical
      prediction (\ref{def:mass_theor}) as dashed line, and
     (c) the parameters $Z$ and $Z_{phot}$
      (using (\ref{def:anomalous_fit},\ref{def:regular_fit}));
     (d) the total monopole density and the monopole density in
     charged clusters with the theoretical
     prediction (\ref{def:density_theor}).
          }
  \label{fig:fit_zeroT}
\end{figure}
Fig.~\ref{fig:fit_zeroT}(a) shows the anomalous dimensions
$\alpha$ first increasing in the low--$\beta$ region. The
anomalous dimension for the {\it angle}-- and for the
{\it sine}--propagator behave quite similar to each other (the dimension
for the {\it angle}--propagator is a bit smaller). As $\beta$ gets larger
than $\beta \sim 1.5$ the anomalous dimensions start to descent towards
zero. This indicates that the anomalous dimension is not only
a function of the monopole density which decreases monotonically
with growing $\beta$ for all values of the coupling. The (cluster) structure
of the monopole configurations may play a significant role for
$\alpha$.

In Fig.~\ref{fig:fit_zeroT}(b) the mass parameters $m$ are presented
for the {\it angle}-- and the {\it sine}--propagator
according to (\ref{def:anomalous_fit}) as function of $\beta$.
Both masses are almost equivalent.
For the mass there exists a theoretical prediction due to
Polyakov~\cite{Polyakov},
\beqn
m_{th}(\beta) = 2 \pi \sqrt{2 \beta} \, \exp\Bigl\{ \pi^2
\beta_V(\beta) \, \Delta^{-1}(0) \Bigr\}\,,
\label{def:mass_theor}
\eeqn
where $\beta_V$ is the Villain coupling constant
\beqn
  \beta_V(\beta) = {\Biggl[2 \log
  \Biggl(\frac{I_0(\beta)}{I_1(\beta)}\Biggr)\Biggr]}^{-1} \, .
  \label{def:beta_V}
\eeqn
$I_0(\beta)$ and $I_1(\beta)$ are the modified Bessel functions
and $\beta$ is the Wilson action coupling constant appearing in
(\ref{def:action}).
The prediction (\ref{def:mass_theor}) is valid for a dilute monopole
gas. The agreement between the two data sets and the theoretical curve
is very good.
The small deviation at lowest $\beta$ can be attributed to the violation of
the dilute gas approximation.

Fig.~\ref{fig:fit_zeroT}(c) shows $Z(\beta)$ and $Z^{phot}(\beta)$
for the two definitions. We observe that $Z$ tends to unity at
large $\beta$, whereas $Z^{phot}(\beta) \approx 1$ for all
$\beta$. The strong deviation of $Z$ from unity at small $\beta$
can be interpreted as a field renormalization by monopoles.

The simplest quantity characterizing the monopoles is the monopole
density, $\rho = \sum_c j_c \slash |\Lambda|$, where the monopole
charge $j$ is defined in Eq.~\eq{def:form_theta_decomposition_2}.
Note, however, that a general monopole ensemble may contain
lattice artifacts which at zero temperature are realized in the
form of ultraviolet monopole-antimonopole pairs. Following
Ref.~\cite{CISPaper1} we remove these lattice artifacts using a
cluster analysis. For our purposes, clusters are defined as
connected groups of monopoles and anti--monopoles, where each
object is separated from at least one neighbor belonging to the
same cluster by a distance less or equal to $R_{\mathrm{max}}$.
We use $R^2_{\mathrm{max}}=3~a^2$ which means that neighboring
monopole cubes should share at least one single corner. The
cluster is called charged if the total charge of its constituent
monopoles is non-zero. This includes isolated
monopoles and anti-monopoles.

In Fig.~\ref{fig:fit_zeroT}(d) we plot the
measured
total monopole density
and the density of monopoles residing in charged
(physical) clusters. The charged fraction is
well
described by the theoretical formula for the monopole
density,
\beqn
  \rho(\beta) = 2 \exp\Bigl\{ - 2 \pi^2 \beta_V(\beta) \,
  \Delta^{-1}(0) \Bigr\}\,.
  \label{def:density_theor}
\eeqn
which is a lattice version~\cite{CISPaper1} of the Polyakov
formula~\cite{Polyakov}. According to Fig.~\ref{fig:fit_zeroT}
both the
monopole density, Debye mass and the deviation of the
coupling  $Z$ from unity are descending functions vanishing in the
limit $\beta \to \infty$.

The contact terms contained in $D^{sin}$ and $D^{ang}$ are not
shown here. The photon part of both $D^{sin}$ and $D^{ang}$
vanishes perfectly. The full propagator in both cases contains
contact terms, $C^{sin}$ and $C^{ang}$, which deviate from zero
for smaller $\beta$, whereas always $C^{sin}(\beta) \ll
C^{ang}(\beta)$.

The discussion of Figs. \ref{fig:momdep_sin_propagator}
and \ref{fig:momdep_ang_propagator} and of the $\beta$ dependence
of the fit parameters in Fig.~\ref{fig:fit_zeroT} was based on the
zero-temperature propagators (and its components)
obtained throughout with $N_G=20$ Gribov gauge copies.
The dependence on the number of gauge copies was investigated
carefully for the case of the {\it angle}--propagator. In the result
of this study, the default choice of $N_G=20$ for the gauge--fixing
procedure at $T=0$ was established.

\begin{figure}[!htb]
  \begin{center}
  \begin{tabular}{cc}
    \epsfxsize=6.0cm \epsffile{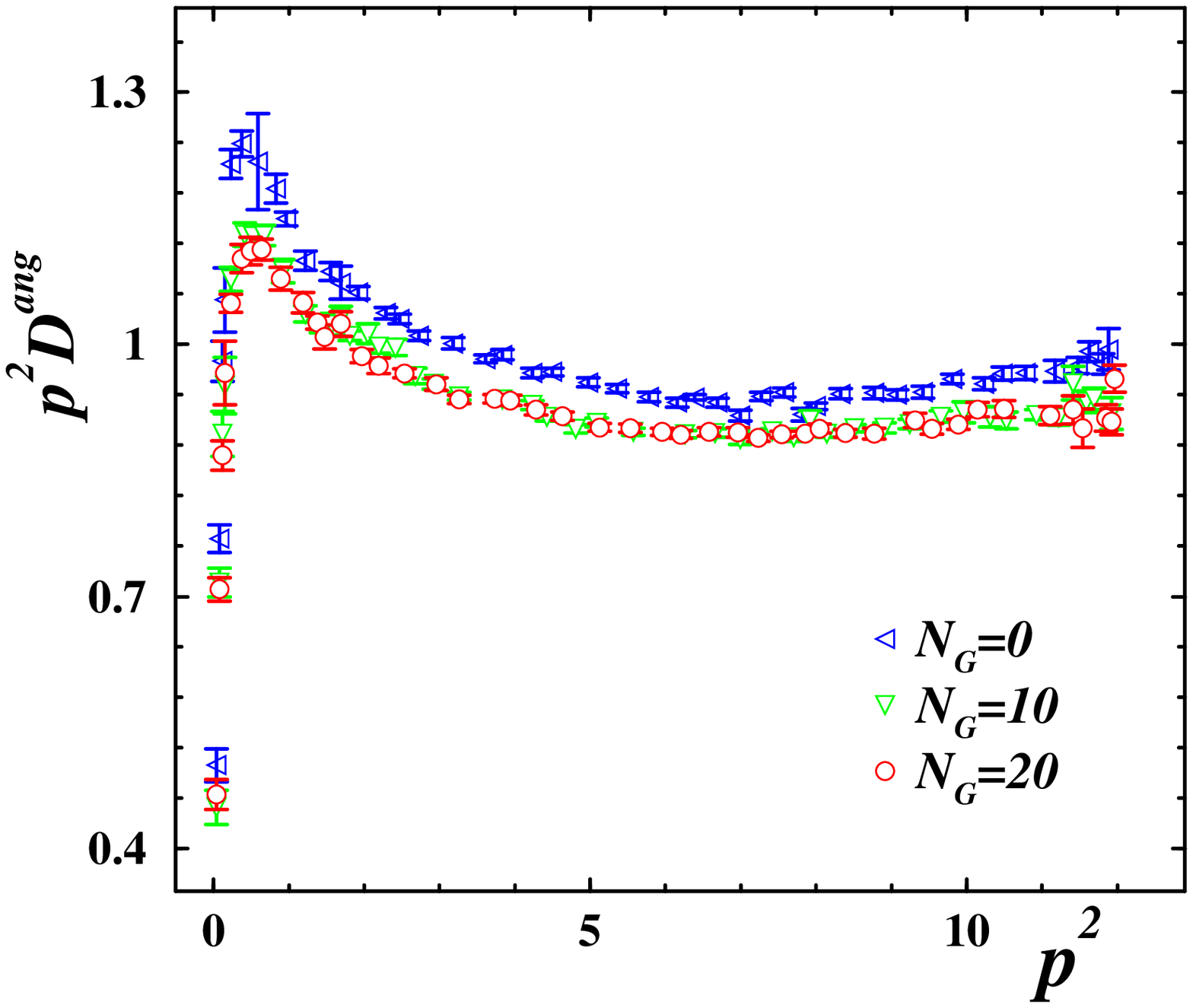}    & \hspace{5mm}
    \epsfxsize=6.0cm \epsffile{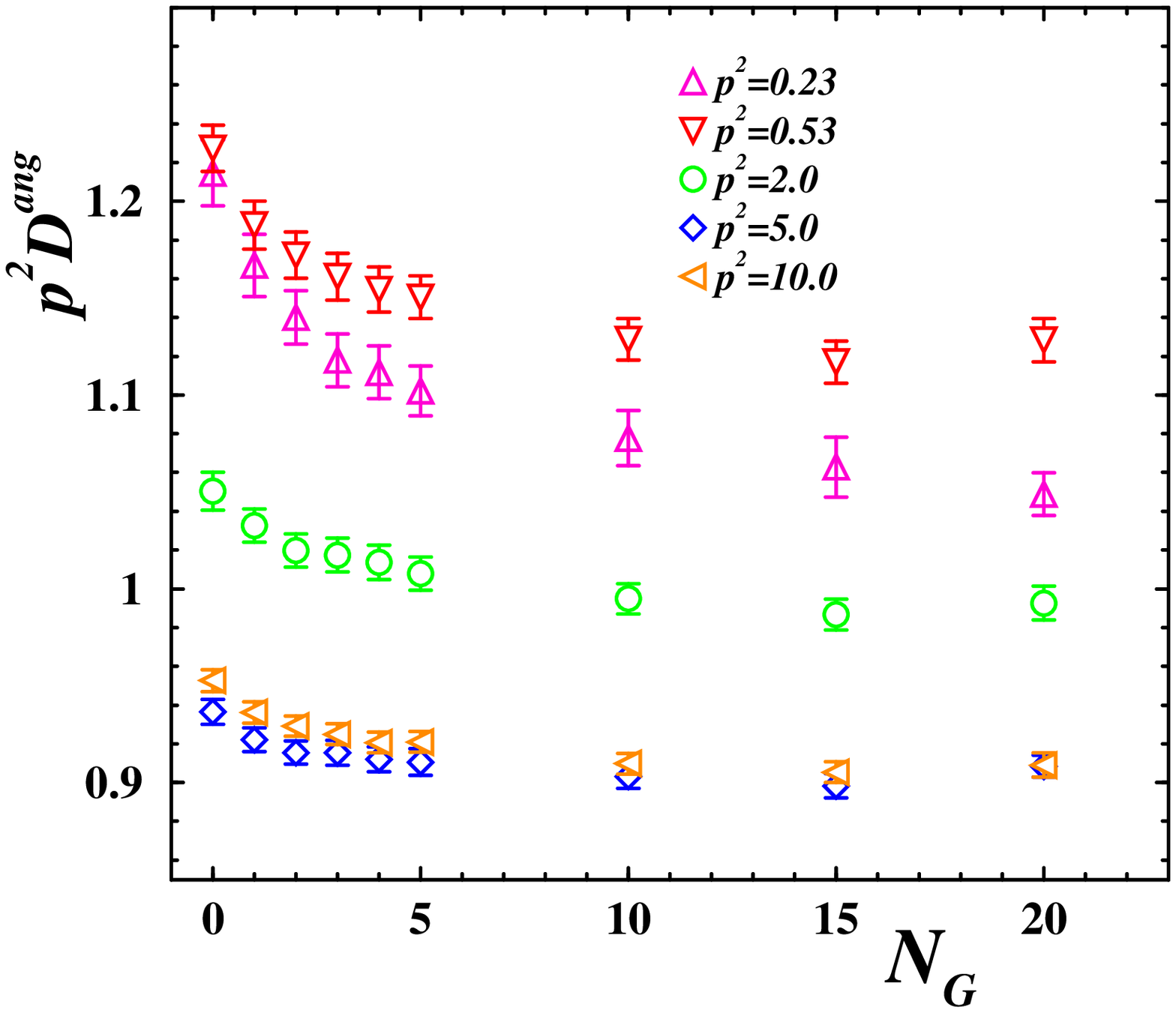} \\
 (a)  & \hspace{5mm}(b)
 \vspace{3mm} \\
    \epsfxsize=6.0cm \epsffile{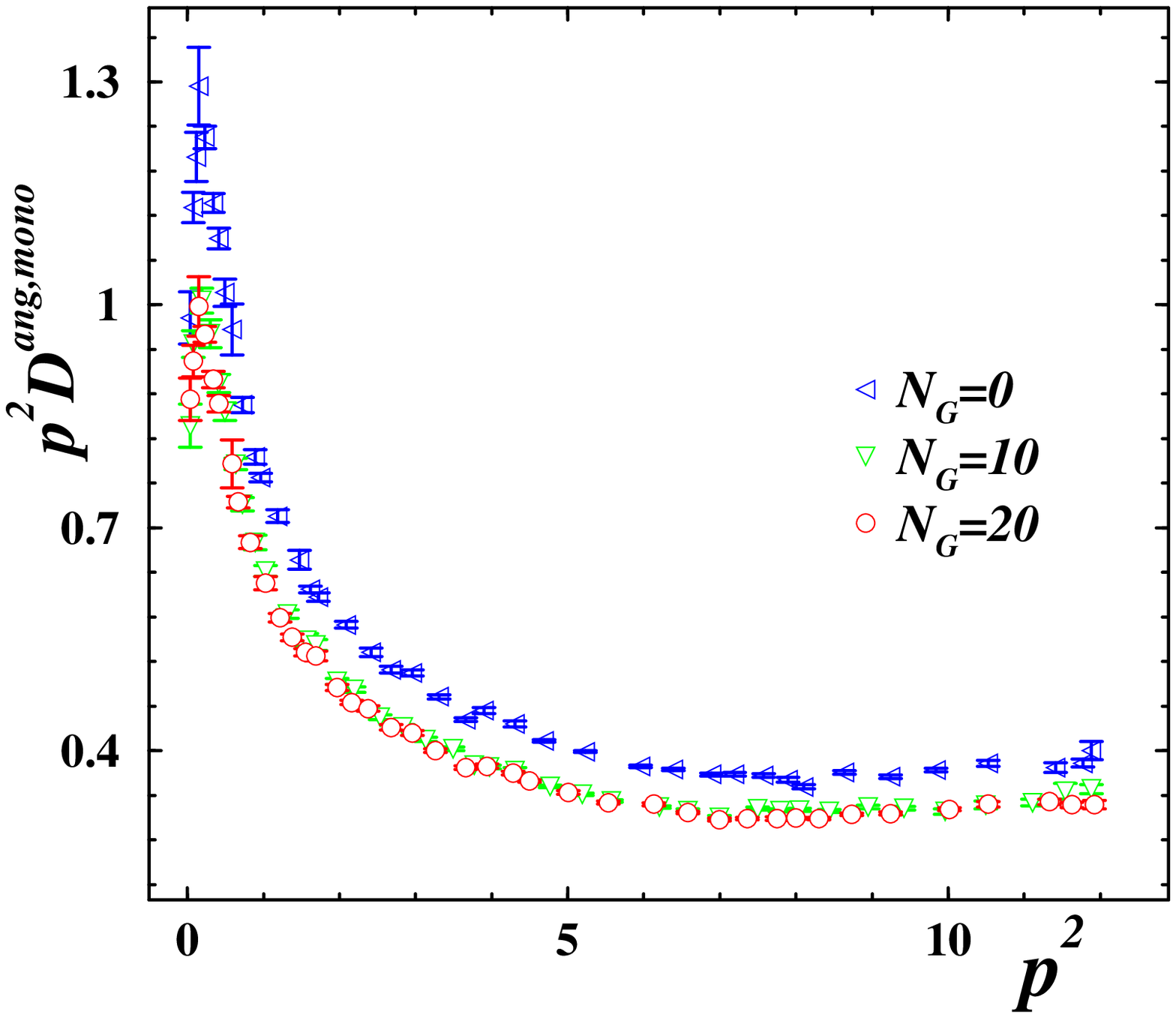}    &\hspace{5mm}
    \epsfxsize=6.0cm \epsffile{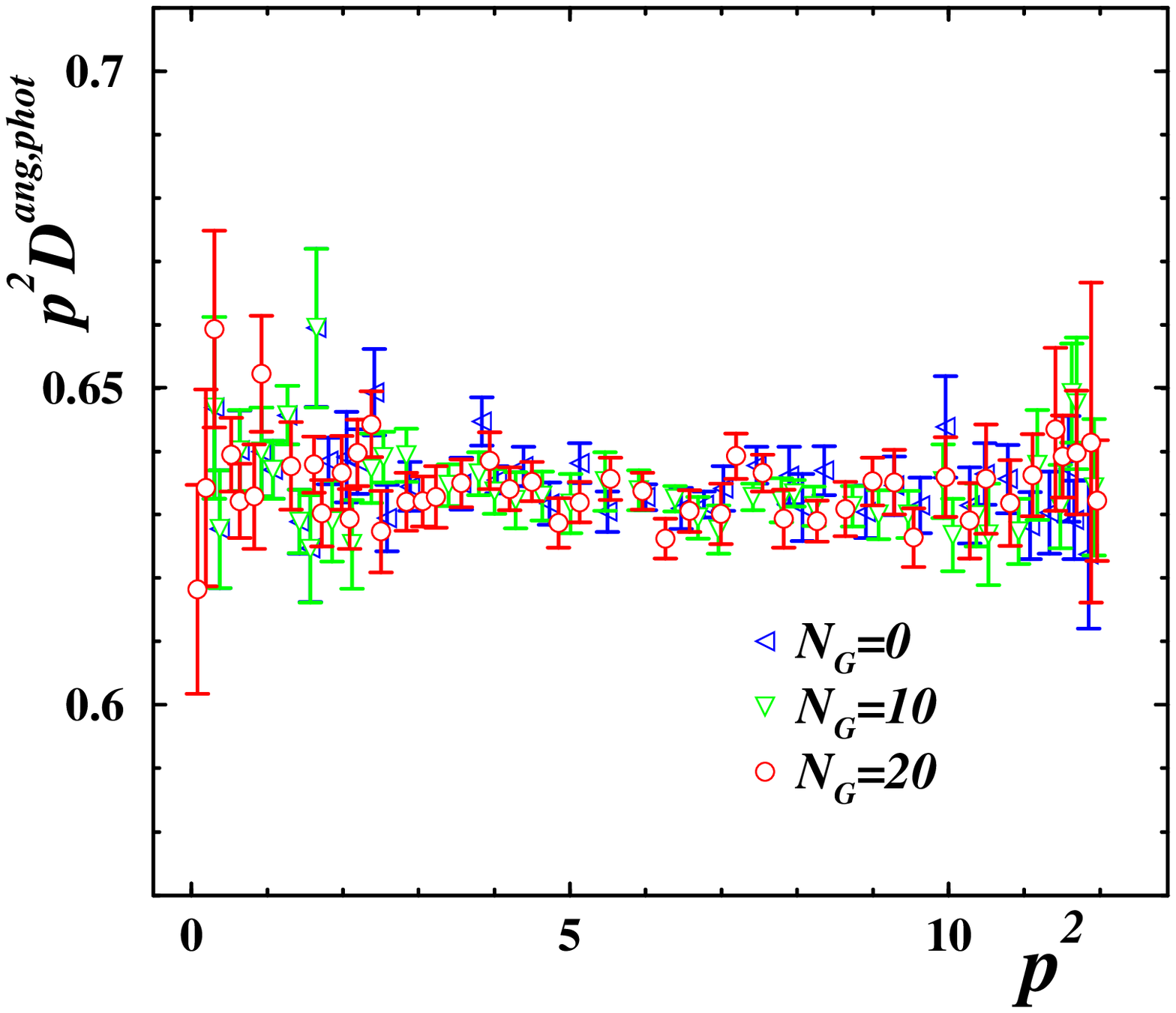}    \\
  (c) & \hspace{5mm} (d)
  \end{tabular}
  \end{center}
  \vspace{-5mm}
  \caption{The dependence of the zero--temperature {\it angle}--propagator on
  $N_G$ for a $32^3$ lattice (at $\beta=2.0$ as an example) :
  (a) the full transverse propagator for $N_G=0,10,20$
  in the full momentum region;
  (b) its behaviour as function of $N_G$ at five selected momenta;
  (c) the same for the singular (mono) part of the transverse propagator;
  (d) the same for the regular (phot) contribution to the transverse
  propagator.
  Again for better presentation part of the data points are not plotted.
  }
  \label{fig:gaugedep_ang_propagator}
\end{figure}

In Fig.~\ref{fig:gaugedep_ang_propagator}
we show different
aspects of the approach to the $N_G \to \infty$ limit, for
$\beta=2.0$ as an example. Fig.~\ref{fig:gaugedep_ang_propagator}(a)
demonstrates that the transverse propagator evaluated with 10
or 20 gauge copies is essentially the same, but that the naive
evaluation (with $N_G=0$) would clearly overestimate the propagator
over the whole momentum range. Fig.~\ref{fig:gaugedep_ang_propagator}(b)
shows this in more detail for five selected momenta. It
becomes clear that the dependence is strongest in the region of
small momenta, in particular the region {\it below} the peak.
The dependence is strong for the singular part presented in
Fig.~\ref{fig:gaugedep_ang_propagator}(c).
Again, there is almost no change between $N_G=10$ and 20.
As can be seen from Fig.~\ref{fig:gaugedep_ang_propagator}(d), there is
almost no $N_G$ dependence in the photon part of the transverse
propagator.

\begin{figure}[!htb]
  \begin{center}
  \begin{tabular}{ccc}
  \epsfxsize=5.0cm \epsffile{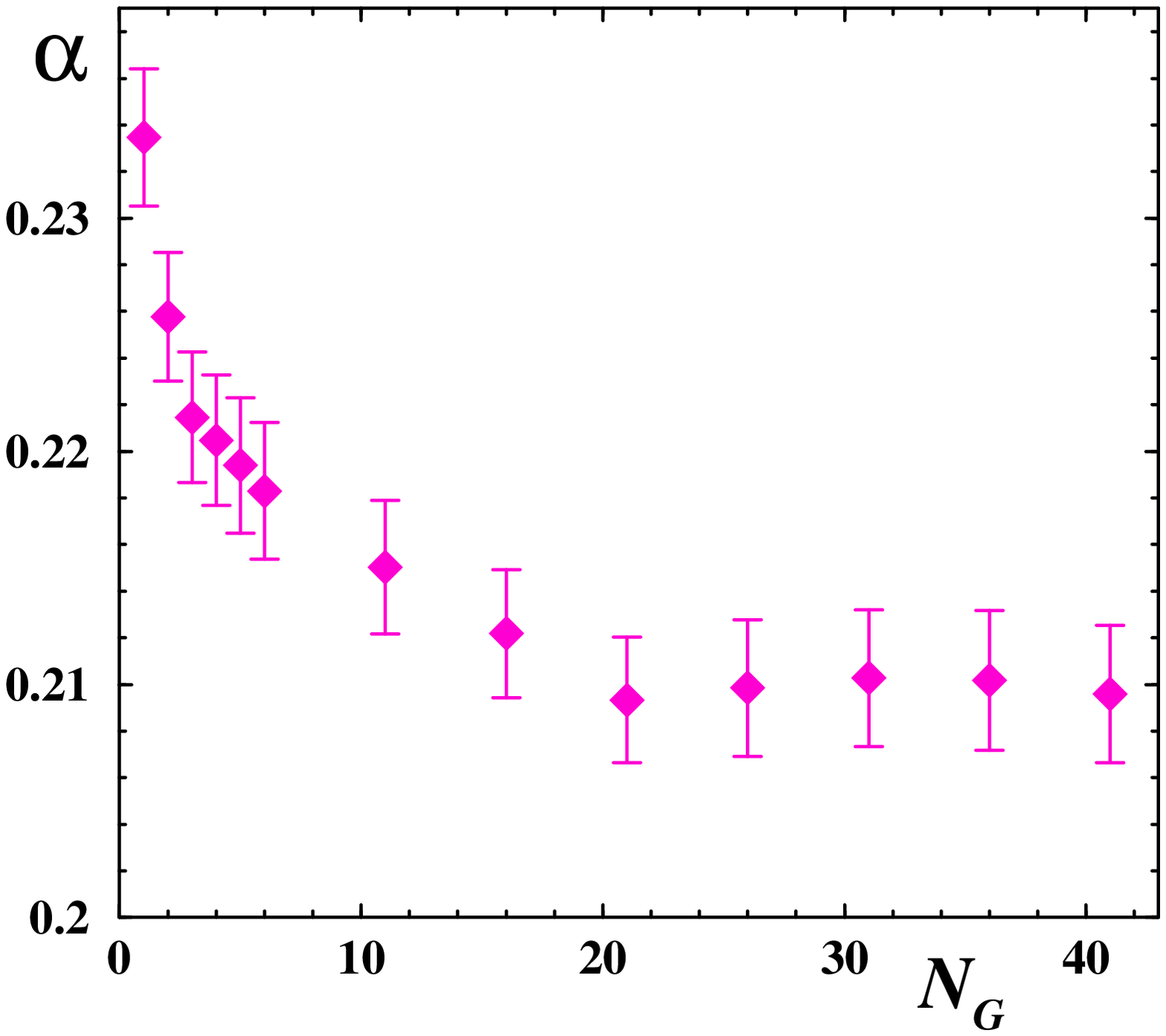} & \hspace{4mm}
  \epsfxsize=5.0cm \epsffile{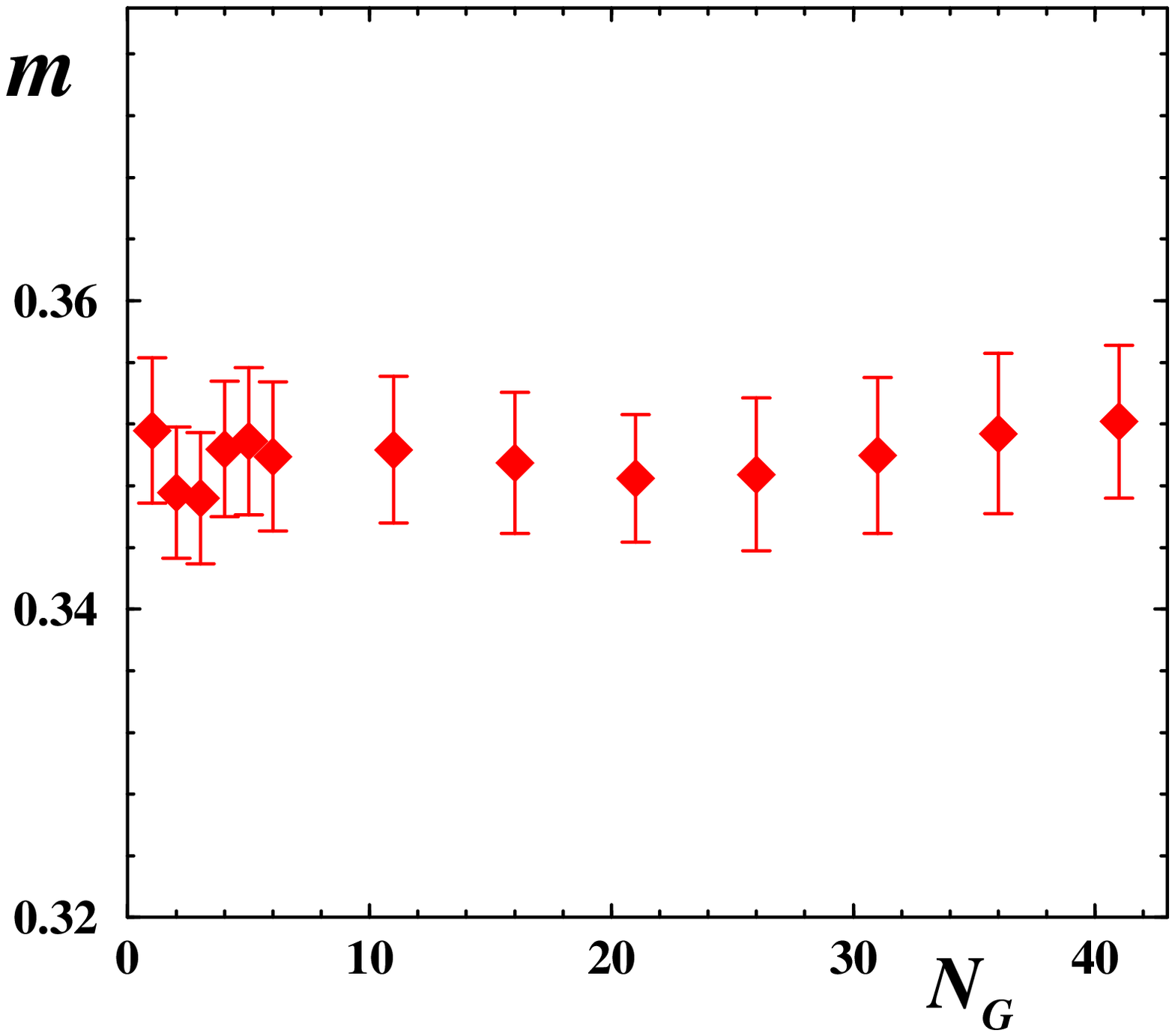} & \hspace{4mm}
  \epsfxsize=5.2cm \epsffile{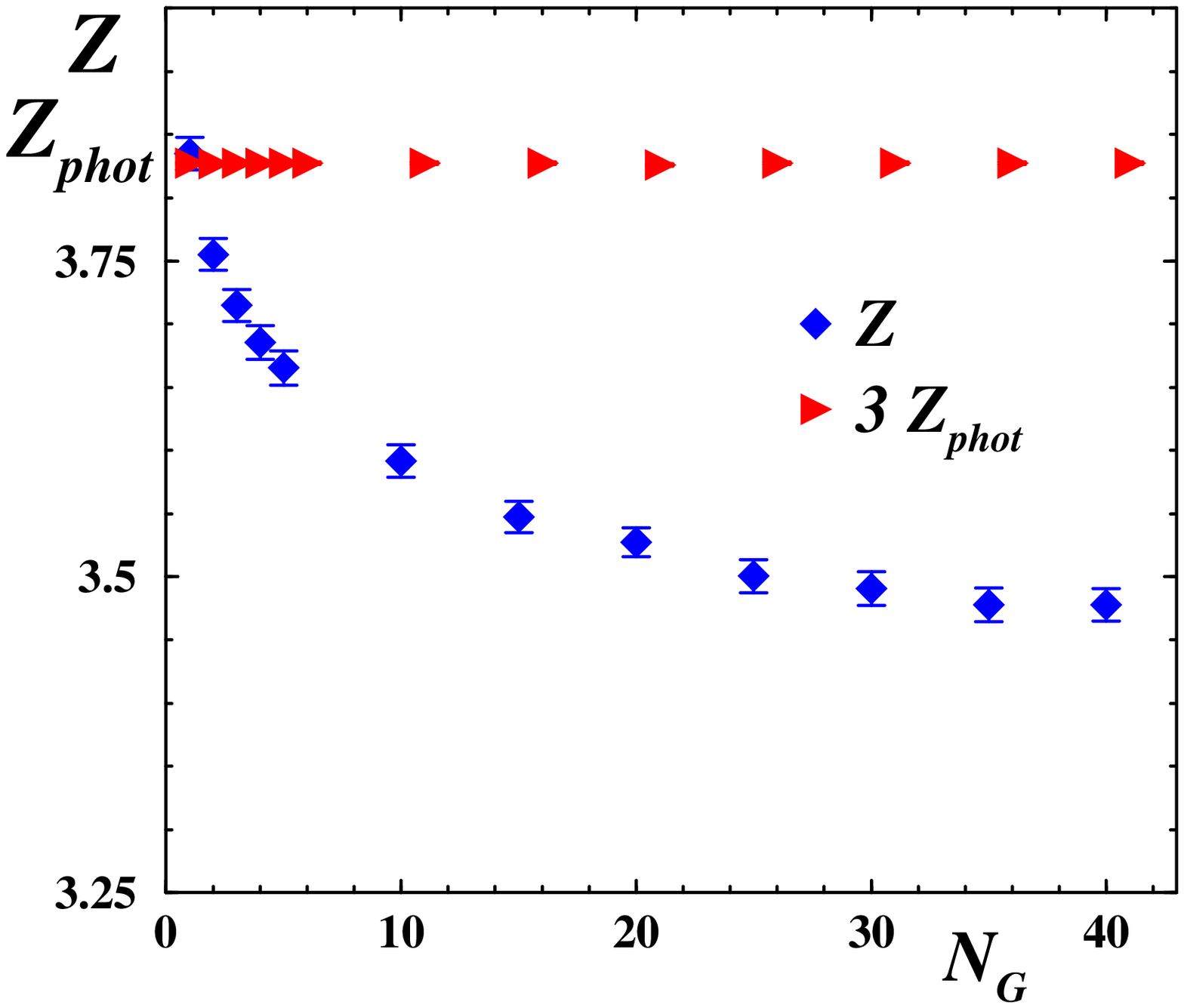} \\
  (a) & \hspace{4mm}(b) & \hspace{4mm}(c)
  \end{tabular}
  \end{center}
  \vspace{-5mm}
  \caption{The best fit parameters for $D^{ang}$ at zero temperature
  as a function of $N_G$ (for $\beta = 2.0$ as an example):
  (a) anomalous dimension $\alpha$,
  (b) mass parameter $m$ and
  (c) renormalization constants $Z$ for $D^{ang}$ and $D^{ang,phot}$.
  }
  \label{fig:gaugedep_fit_parameters}
\end{figure}

We present the resulting dependence of the fit parameters on $N_G$
in Fig.~\ref{fig:gaugedep_fit_parameters},
again for $\beta=2.0$.
The anomalous dimension $\alpha$ shown in
Fig.~\ref{fig:gaugedep_fit_parameters}(a) drops within 10 \% which
indicates that the anomalous dimension is sensitive with respect
to the minimization of the Dirac strings that is achieved by
better and better Gribov copies.
The mass presented in Fig.~\ref{fig:gaugedep_fit_parameters}(b)
does not change with $N_G$ which confirms that it is mainly
determined by the monopole mass ({\it i.e.} density).~\footnote{Let us
recall that the monopole positions in a given configuration are
gauge independent.}
Fig.~\ref{fig:gaugedep_fit_parameters}(c) shows $Z$ for the full transverse
propagator and $Z^{phot}$ for the photon part.
It is not surprising that the parameter associated with the photon part do
not change.
The parameter describing the full transverse propagator decreases,
again by $\approx 10 \%$ within $N_G < 20$.
One can also observe a slight dependence of $Z$ on $N_G$ for
$N_G>20$. However, the dependence is indeed very small (about 1\%)
and not essential for our qualitative discussion.

{}From the comparison of the {\it sine}--propagator and the
{\it angle}--propagator at zero temperature we conclude that the fits of
the transverse part give more or less the same parameters, with a
$\beta$ dependence (if appropriate) which is in accordance with
the monopole density. Strict transversality itself is guaranteed only
in the case of the {\it sine}--propagator. In the case of the
{\it angle}--definition an appropriate transverse part has to be
extracted by projection
(\ref{def:F_project_2},\ref{def:D_project_2}).


\section{The finite-temperature propagator in Landau gauge}
\label{sec:finite_temperature}

In this Section we report our investigation of the properties of
the gauge boson propagator at finite temperatures. With respect to
the distinguished direction $\mu=3$, the propagator $D$ can be
separated into transverse and longitudinal components (see
Section~\ref{sec:model_and_propagator} for the definitions)
denoted as $D_T$ and $D_L$, respectively. We are working on the
lattice $32^2 \times 8$, in line with Ref.~\cite{CISLetter}, where
only the $D_L$ component of the {\it angle}--type propagator
(with $p_3=0$, $N_G=20$ and a limited statistics of $500$ measurements)
was studied.

The transverse component of the propagator, $D_T$, describes the spatial
degrees of freedom while the longitudinal component, $D_L$,
contains gauge fields in Landau gauge in both temporal and spatial
directions. The finite temperature propagator data are analyzed again for
$p_3=0$, as a function of ${\mathbf p}^2$.
In that case $D_L$ is constructed only from temporal degrees of freedom
which, in particular, are responsible for the confinement phenomena.
We have fitted the data for both components of the propagator using the fit
function~\eq{def:anomalous_fit} invented first in
Ref.~\cite{CISLetter} to describe $D_{33}$.

\begin{figure}[!htb]
  \begin{center}
  \begin{tabular}{cc}
  \epsfxsize=6.0cm\epsffile{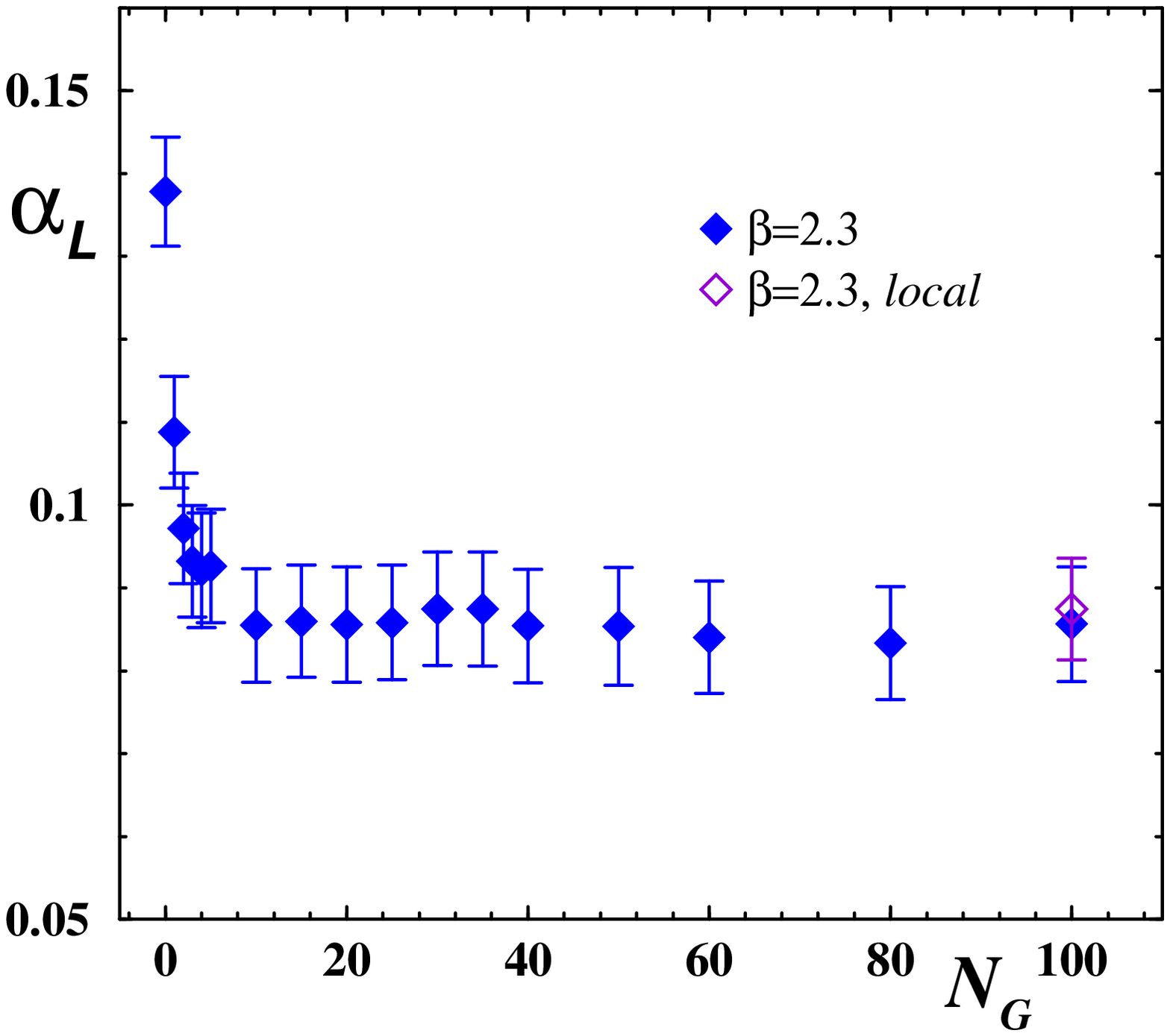} & \hspace{5mm}
  \epsfxsize=6.0cm\epsffile{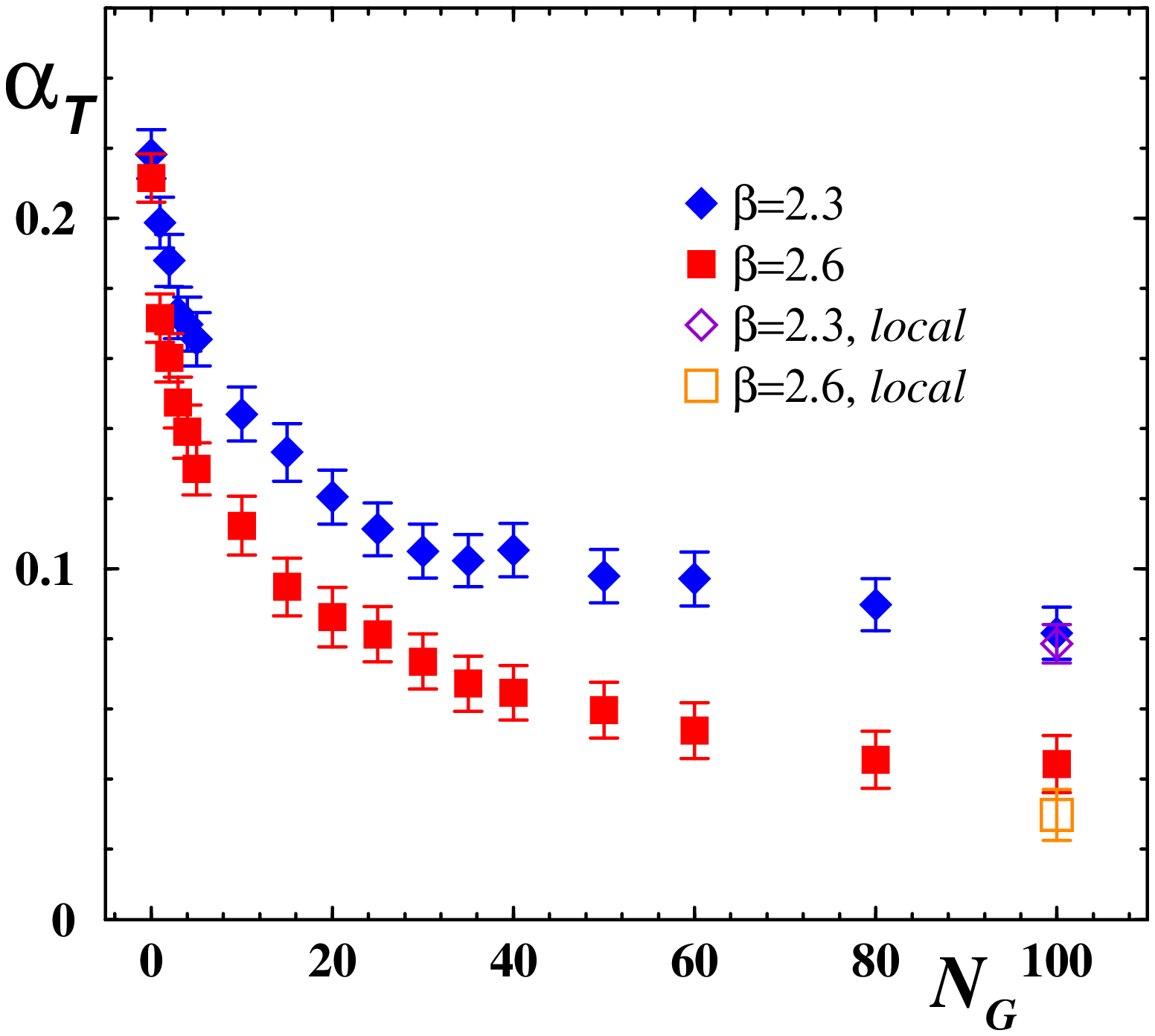} \\
  (a) & \hspace{5mm} (b) \\
  \\
  \epsfxsize=6.0cm\epsffile{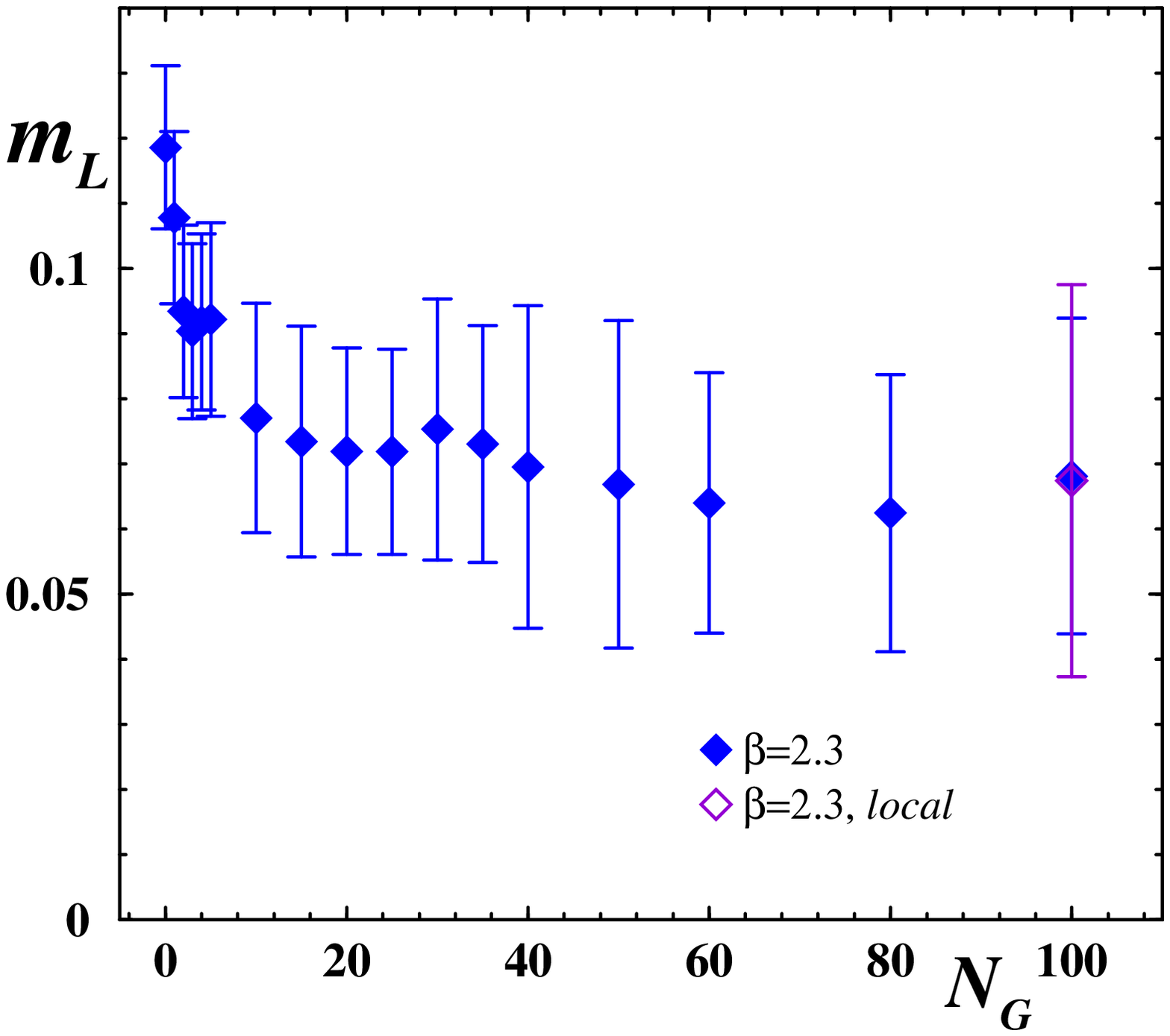}  & \hspace{5mm}
  \epsfxsize=6.1cm\epsffile{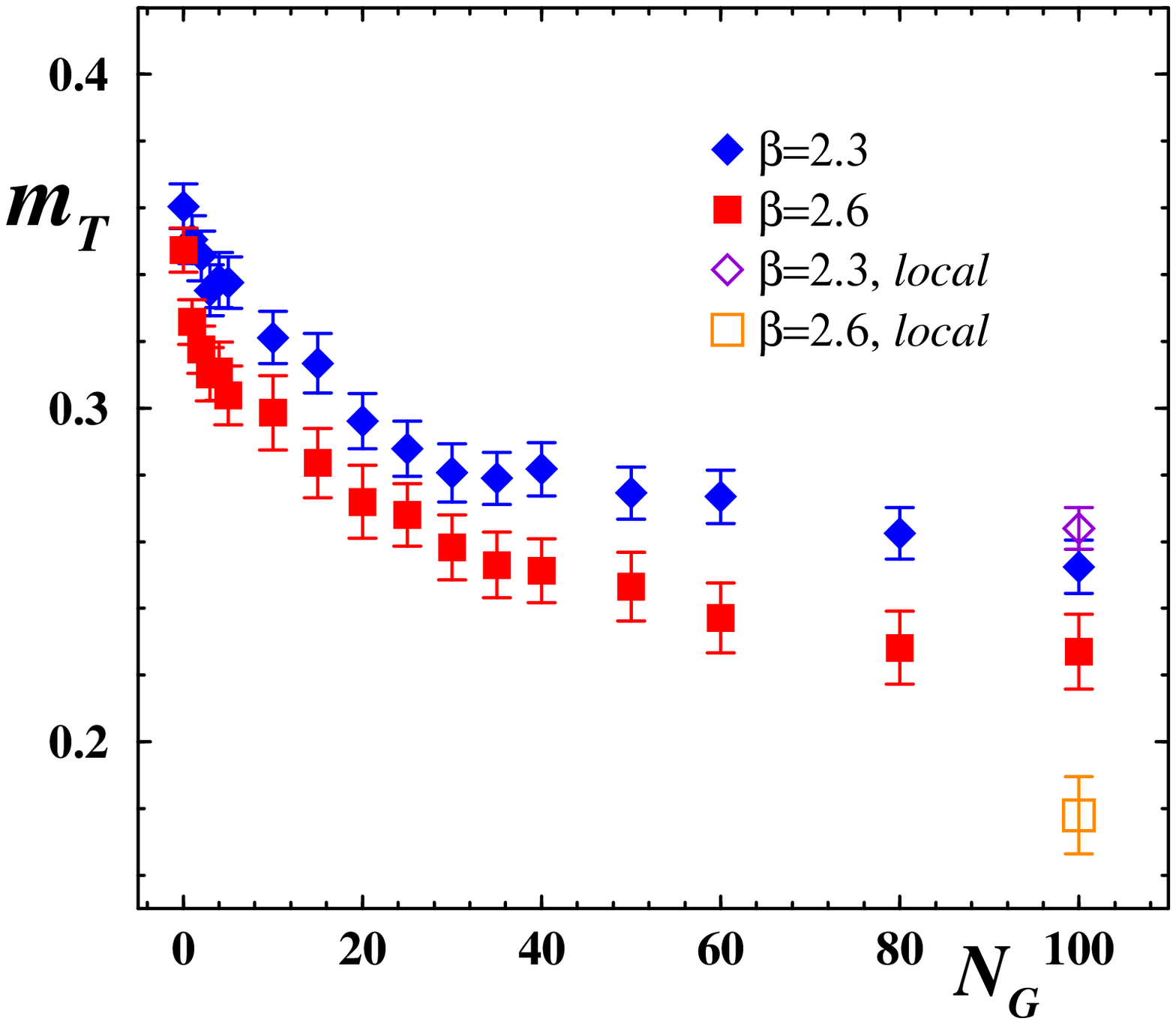}  \\
  (c) & \hspace{5mm} (d) \\
  \\
  \epsfxsize=6.0cm\epsffile{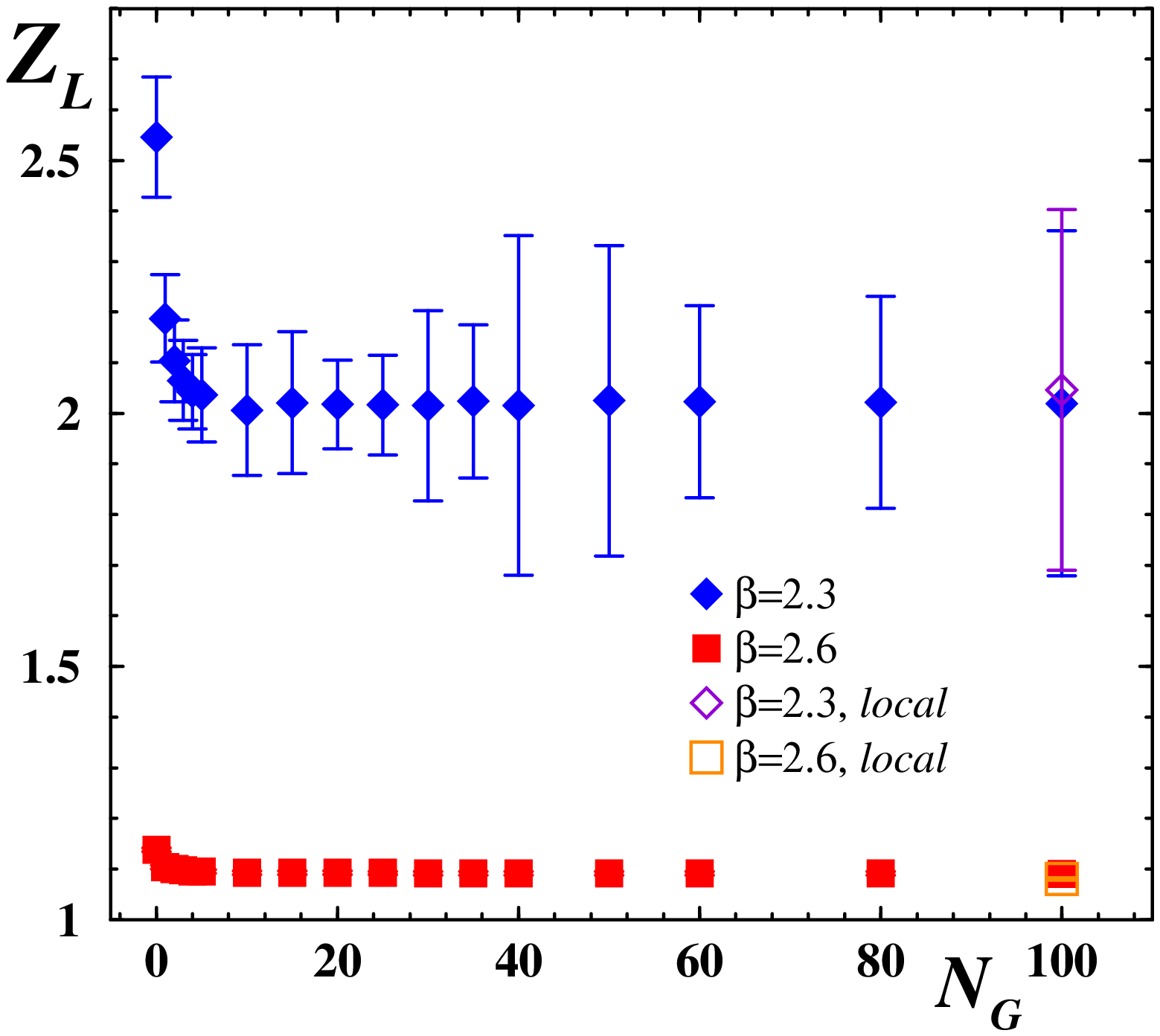}     & \hspace{5mm}
  \epsfxsize=6.0cm\epsffile{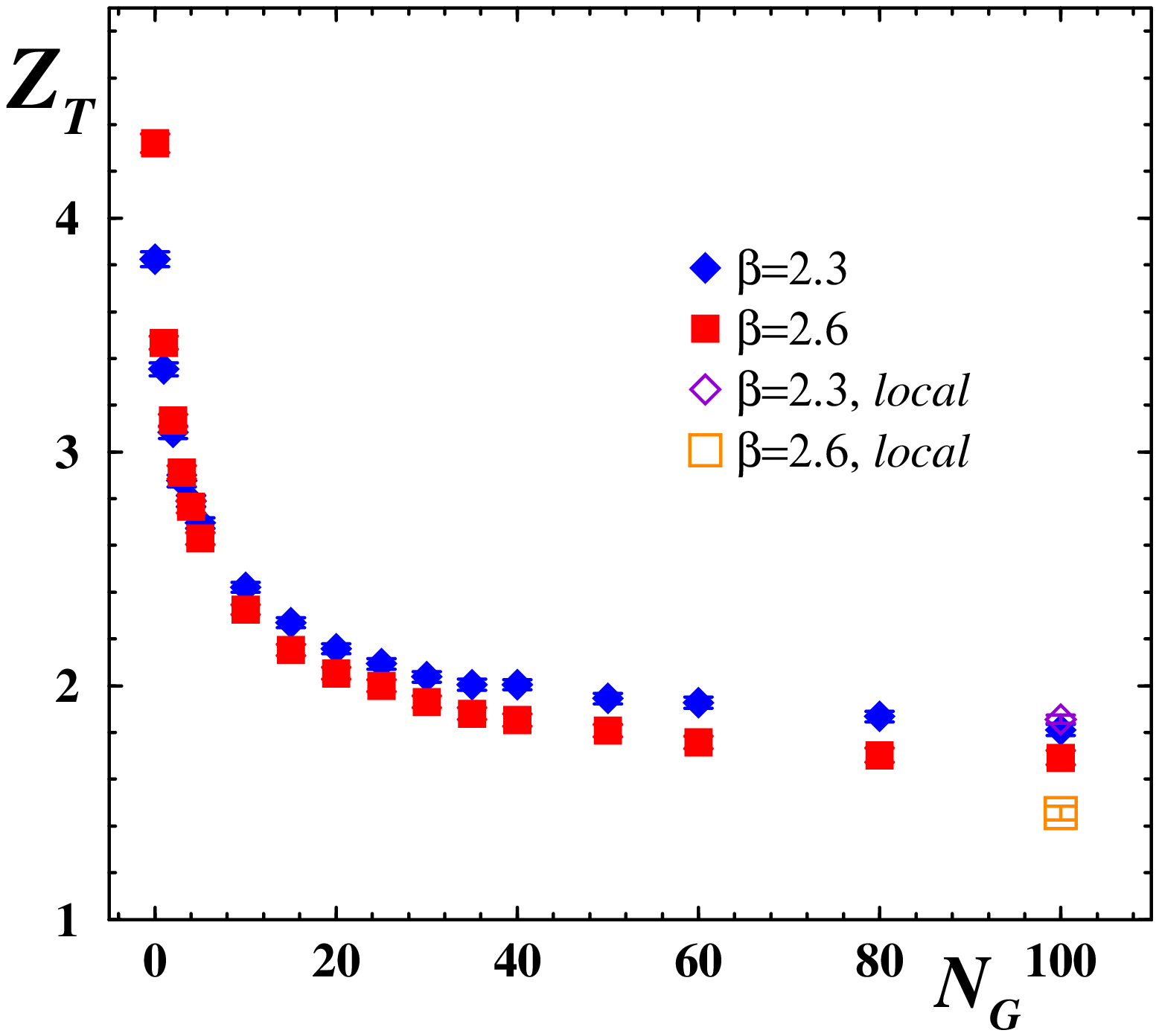} \\
  (e) & \hspace{5mm} (f)
  \end{tabular}
  \end{center}
  \vspace{-5mm}
  \caption{ The Gribov copy dependence of the best fit parameters
  for the $D_L$ and $D_T$ propagators:
  the anomalous dimension $\alpha$ (a,b),
  the mass parameter $m$ (c,d) and
  the parameters $Z$ (e,f).
  The label ``local'' corresponds to measurements in Monte Carlo cycles
  without global updates and with $N_G=100$ extra Gribov copies.
  }
  \label{fig:gribov_nonzeroT}
\end{figure}

First we repeated the investigation of the Gribov copy
dependence of the propagator components $D_L$ and $D_T$,
this time for the {\it sine}-definition of the propagator,
similar to that conducted for the zero--temperature case with
the {\it angle}-definition. The results are
summarized in Fig.~\ref{fig:gribov_nonzeroT}
at $\beta$  values near (below and above) the phase transition.
After a few Gribov copy attempts the longitudinal component
is almost insensitive to the number $N_G$.
This can be seen from the left panel of Fig.~\ref{fig:gribov_nonzeroT}.
The fitting parameters $\alpha_L$, $m_L$, $Z_L$ and $C_L$ (the latter is
not shown here) are rapidly converging and become almost independent
of $N_G$ for $N_G \gtrsim 7$. The results at large number of
Gribov copies are not sensitive to the fact whether we have
suppressed global updates (we made only local updates, see the
label ''local '') or not.

The transverse component, however, is strongly dependent on
the number of Gribov copies as it can be seen from
Fig.~\ref{fig:gribov_nonzeroT}. All fit parameters
$\alpha_T$, $m_T$ and $Z_T$ are descending functions of the
number of Gribov copies $N_G$ with
global updates included. At $N_G=100$ the plateau is not yet reached.
Moreover, the results are sensitive to whether or not
global updates are included, in particular at high values of $\beta$.
The measurements with only local updates lead to significantly lower fit
results. In deconfinement we would expect vanishing $m_T$ and $\alpha_T$),
and $Z_T \to 1$.

The reason for that behavior might be explained as follows:
On one side, the ''best'' gauge functional is realized in {\it gauge--fixed}
configurations without Dirac lines wrapping around the lattice (predominantly
in the short temporal direction).
Such Dirac lines are continuously created and
destroyed by the Monte Carlo process, even if global updates are {\it not}
attempted.
The level of ``noise'' due to wrapping Dirac lines is higher if global
updates are included in the Monte Carlo process, which in general would
improve the ergodicity of the system, but it represents a problem also
if only local updates are used.
The presence of wrapping Dirac strings mimics a finite
$\beta$-independent {\it lattice} mass $m_T$ at larger $\beta$, the value of
which decreases only with increasing temporal extent $L_T$. So in the limit of
vanishing lattice spacing the dimensionful mass would diverge.

This ''Dirac noise'' represents a serious challenge for the gauge fixing
algorithm. The deterministic part (overrelaxing steepest descent method)
described in Section~\ref{subsec:LandauGauge} cannot remove it.
{\it Unbiased} random gauge transformations applied to get new start
configurations for the deterministic search for further Gribov copies
are obviously not effective enough to reduce the ''Dirac noise''.
A simulated annealing Monte Carlo series of random gauge transformations
with the total length of Dirac strings as ''gauge action'' \cite{Kerler}
seems to be more appropriate for selecting new start configurations
for the final steepest descent search.

Having these difficulties in mind, we decided to use in the final measurements
at finite temperature, for $\beta=2.0$ and larger, only local updates before
gauge fixing and to perform $N_G=100$ Gribov copy attempts.
For both {\it sine}-- and {\it angle}--propagator measurements  deep in the
confinement phase (below $\beta=2.0$) we used $N_G=20$ and global updates
where the results of the zero temperature analysis for the Gribov copy
dependence is applicable and the fit parameters of both $D_L$ and $D_T$ have
to agree within accuracy and should be similar to those for the $T=0$
transverse propagator $D$. Nevertheless, we have to admit that the results for
the transverse propagator $D_T$ should be understood only qualitatively.

\begin{figure}[!htb]
  \begin{center}
  \begin{tabular}{cc}
  \epsfxsize=6.cm \epsffile{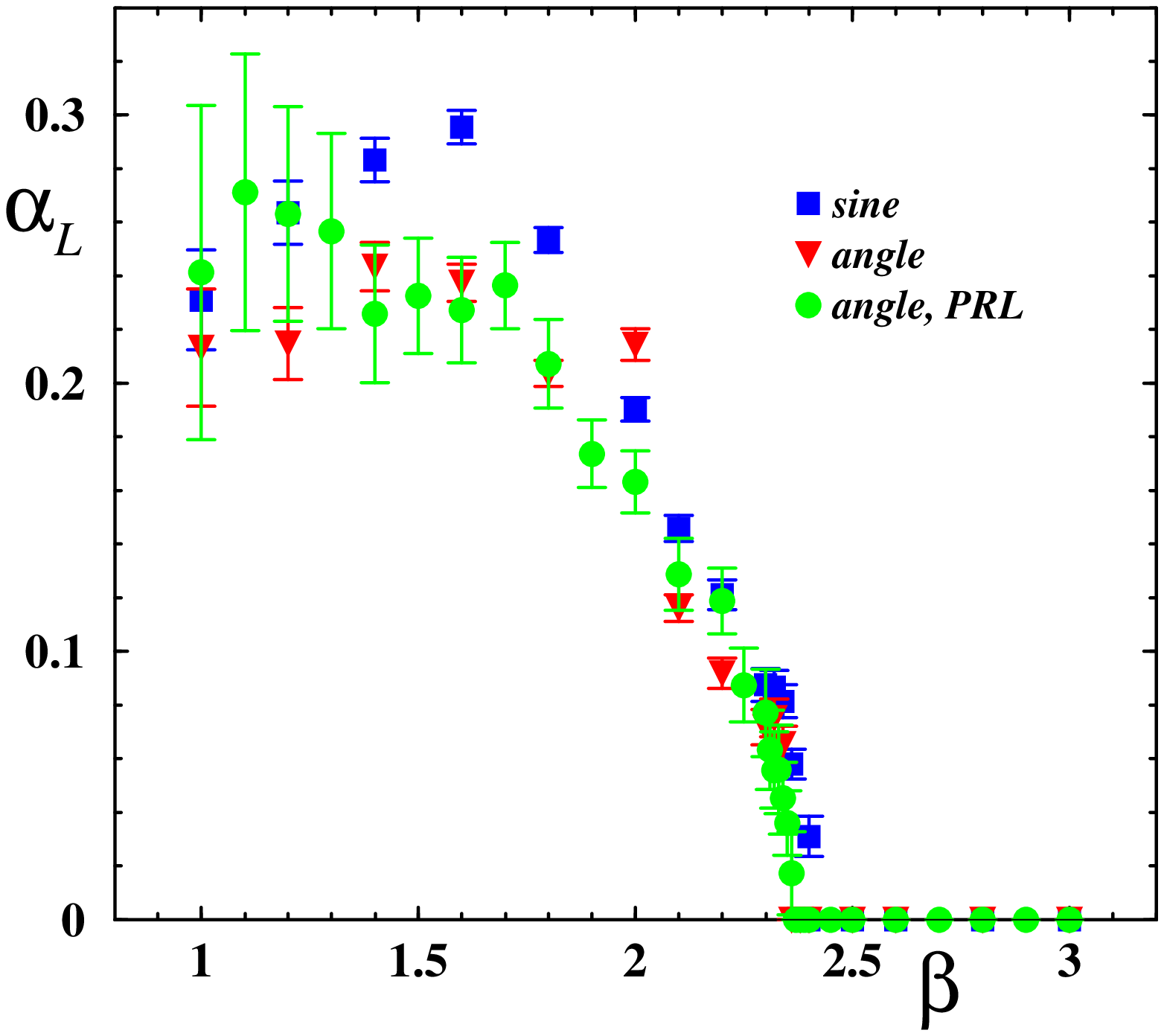} & \hspace{5mm}
   \epsfxsize=6.cm \epsffile{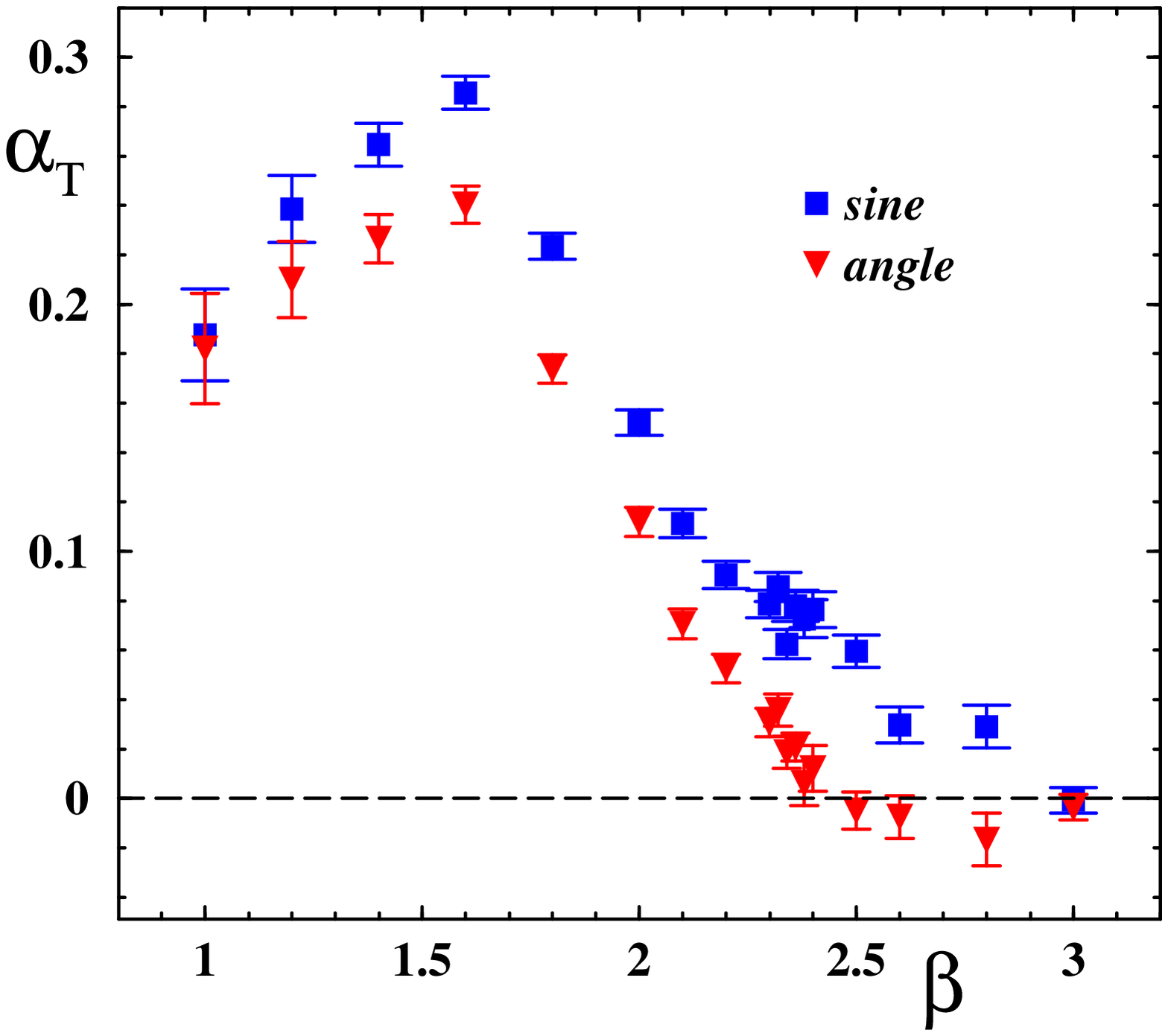} \\
   (a) & \hspace{5mm}(b) \\
   \\
  \epsfxsize=6.cm \epsffile{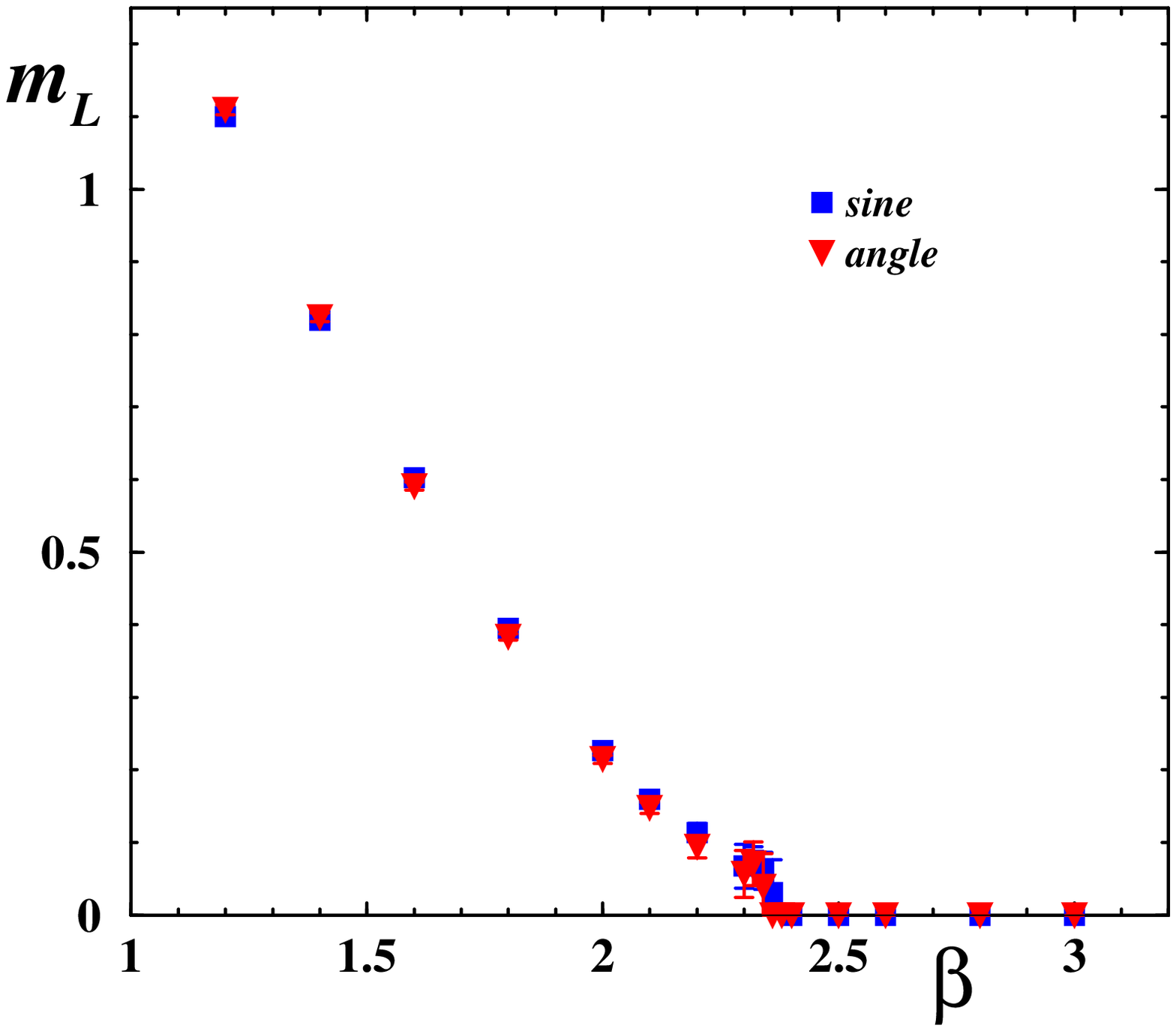} & \hspace{5mm}
  \epsfxsize=6.cm \epsffile{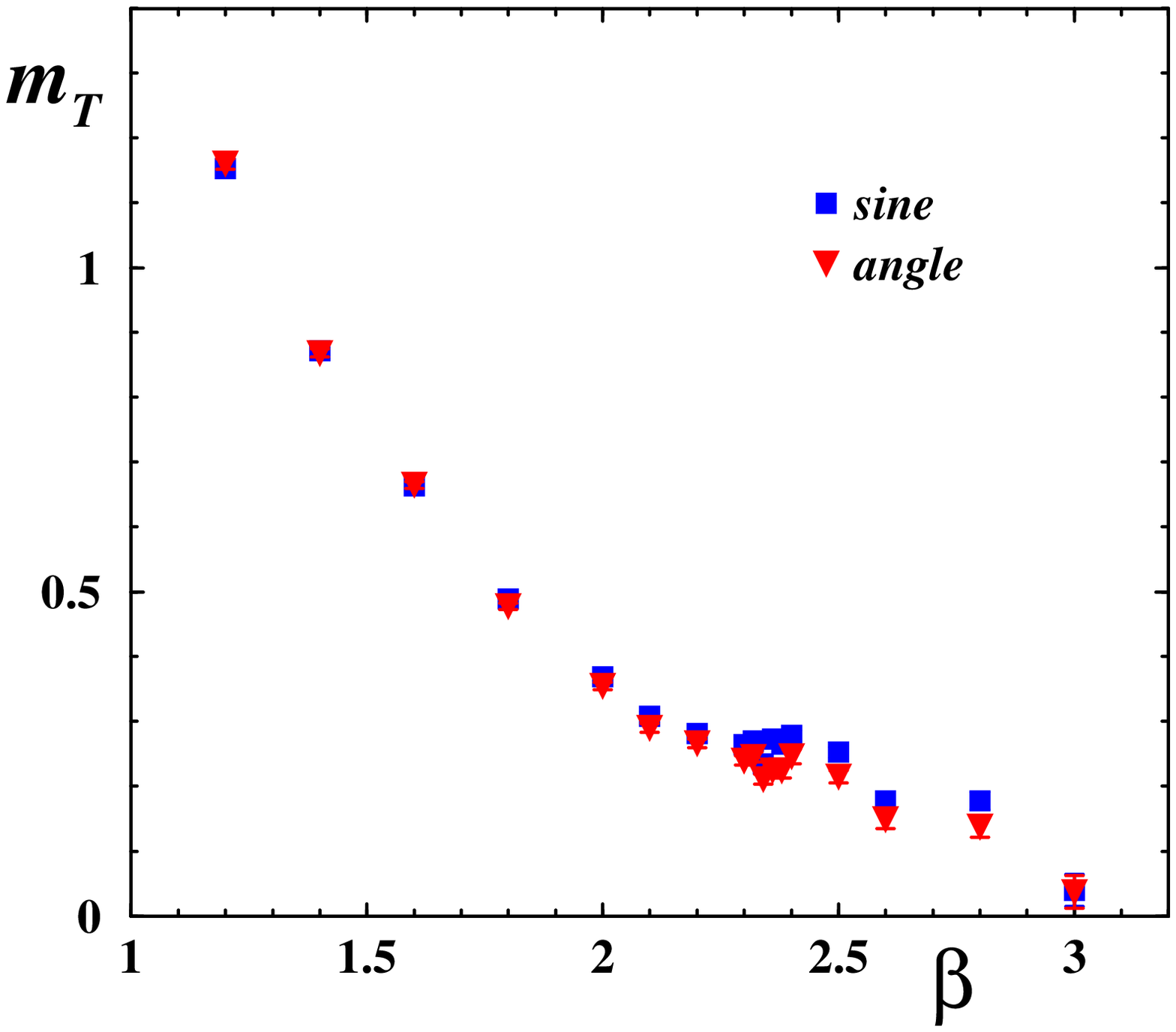} \\
   (c) & \hspace{5mm}(d) \\
  \\
  \epsfxsize=6.cm \epsffile{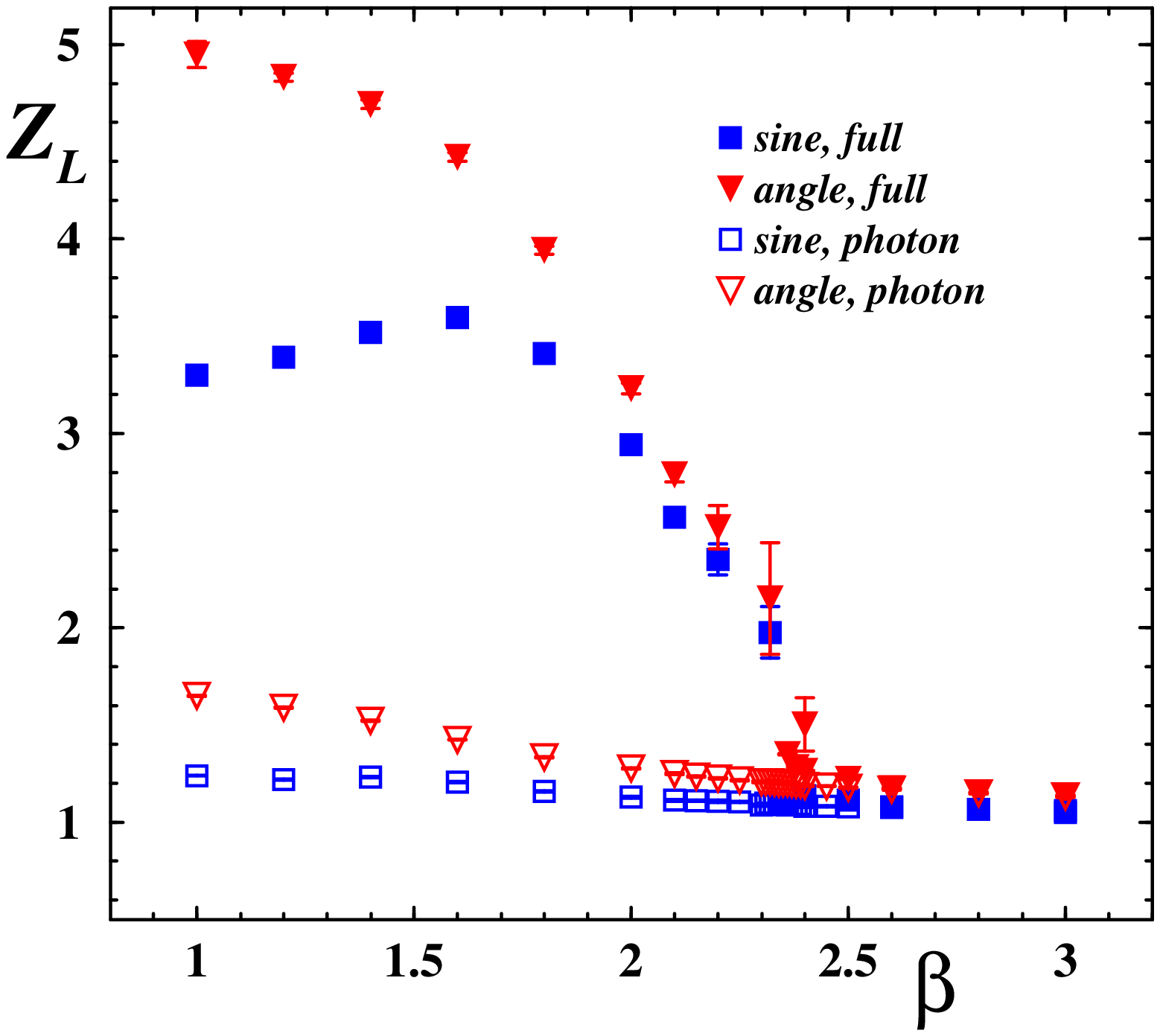} & \hspace{5mm}
  \epsfxsize=6.cm \epsffile{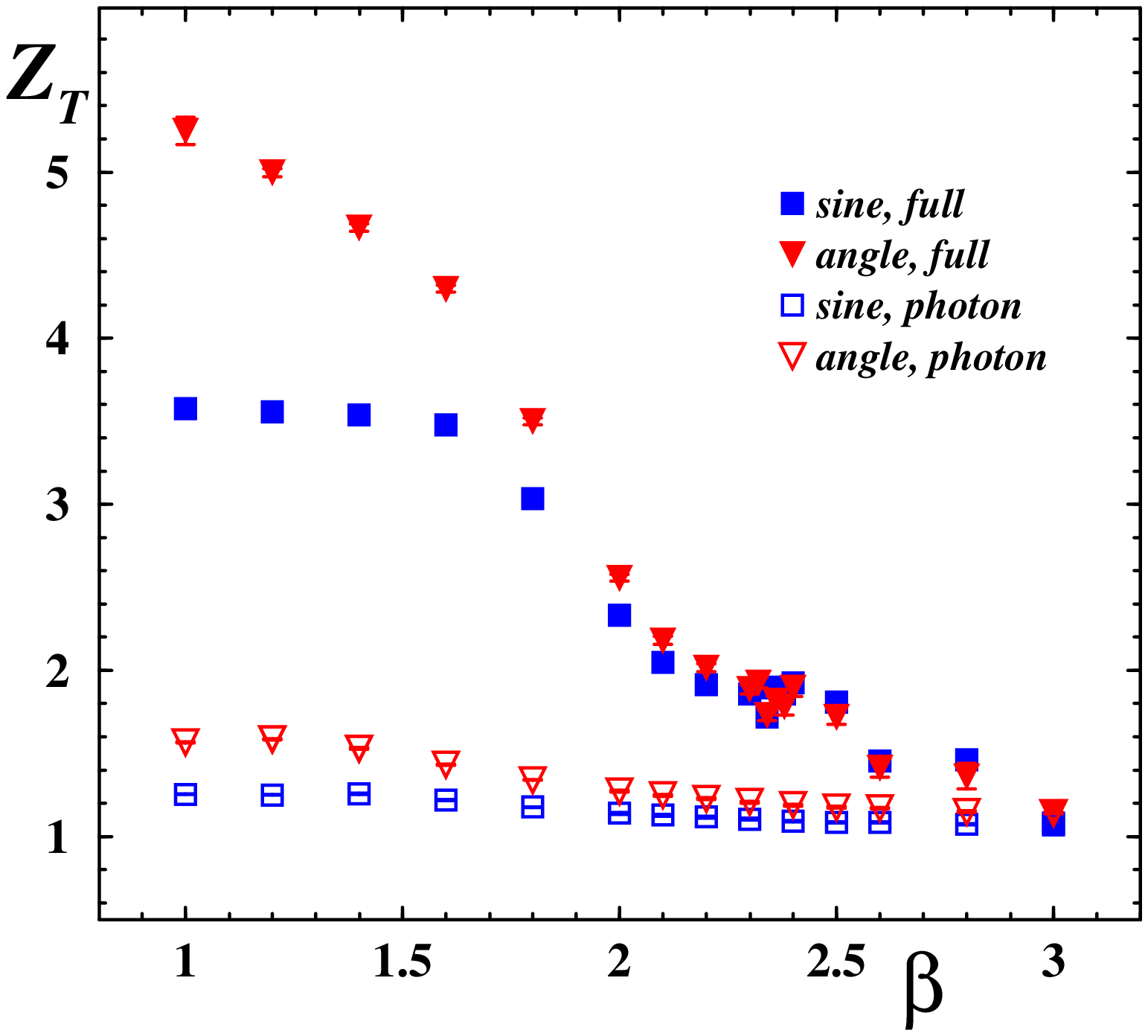} \\
  (e) & \hspace{5mm}(f)
  \end{tabular}
  \end{center}
  \vspace{-5mm}
  \caption{The best fit parameters for the
   non--zero temperature {\it sine}-- and {\it angle}--propagators
   as functions of $\beta$:
   the anomalous dimension $\alpha$ (a,b), the mass parameter $m$ (c,d),
   the parameters $Z$ and $Z_{phot}$ (e,f).
   The left column corresponds to $D_L$, the right to $D_T$.
   The fits of full propagators are done with the help of
   (\ref{def:anomalous_fit}), the photon contribution is fitted by
   (\ref{def:regular_fit}).}
  \label{fig:fit_nonzeroT}
\end{figure}

The results for best fit parameters for $D_L$ and $D_T$ are  presented in
Fig.~\ref{fig:fit_nonzeroT}.
Let us begin with $D_L$. At the critical point
both the anomalous dimension $\alpha_L$ and the mass $m_L$ vanish
while the renormalization parameter $Z_L$ meets the
corresponding parameter for the perturbative photon, $Z_L^{phot}$.
This behaviour is characteristic for both {\it angle} and {\it sine} types of
the propagator and extends our results in
Ref.~\cite{CISLetter}. Note that also here
the mass parameters for {\it sine}-- and
{\it angle}--propagators coincide with each other. However, the
anomalous dimensions for these cases differ slightly from each
other while the renormalization parameters are significantly
different, and the renormalization factor
for the {\it angle}--propagator is bigger than
that for the {\it sine}--propagator. The latter is expected because
$|\sin\theta| \leqslant |\theta|$.

The corresponding quantities for $D_T$ behave differently for the
{\it angle}-- and {\it sine}--definitions of $A_{\mu}$, with the
remarkable exception of the mass parameter. For example, the
anomalous dimension $\alpha_T$ for the {\it angle}--propagator
vanishes in the vicinity of the critical point and beyond, while
the same quantity for the {\it sine}--propagator does not vanish.
We explain this behaviour as due to insufficient gauge fixing as
it can also be guessed from our previous analysis.
The same reason explains the fact that the masses $m_T$ for both
definitions of the photon propagators -- being remarkably similar
-- are not vanishing at the critical point. Finally, for both
definitions of the propagators the renormalization constants $Z_T$
do not approach the corresponding $Z^{phot}_T$ at the critical
point. In order to get a reliable behaviour of $D_T$ part of the
propagator one should drastically increase the number of Gribov
copies used in the gauge fixing. For the time being this is beyond
our computing capabilities.
The situation could be improved using a variant of the mentioned
simulated annealing Monte Carlo series of random gauge transformations in
order  to choose more appropriate initial gauge  transformed configurations
before fixing the gauge.


\section{Conclusions}
\label{sec:conclusions}

We have studied the gauge boson propagator in cQED$_3$ both at
zero and non--zero temperatures. We have found that the
propagators in all cases under investigation can be fitted by
(\ref{def:anomalous_fit}) which is the sum of the massive
propagator with an anomalous dimension plus a contact term.
Similarly to the case of $D_L$ at finite--temperature~\cite{CISLetter}, a
nonvanishing anomalous dimension $\alpha$ is also found at $T=0$.
Moreover, the
fact of existence of the anomalous dimension is not associated
with a particular type of gauge boson propagator. We have
studied {\it angle}-- and {\it sine}--types of propagators and the
corresponding anomalous dimensions are nonvanishing and have a similar
behaviour as the functions of $\beta$.

The existence of the anomalous dimension depends on the presence
of the monopole plasma, but it is not directly proportional to the
monopole density below $\beta=1.5$. In the confinement phase the
monopole plasma is present at any coupling of the system and the
density of monopoles is a monotonously decreasing function of the
lattice coupling $\beta$. A similar behaviour is observed for the
anomalous dimension in the case of the $D_L$ and $D_T$
propagators. The dimension $\alpha_L$ extracted from the $D_L$
component of the propagator is vanishing in the vicinity of the
phase transition for both definitions ({\it angle} and {\it
sine}) of the propagator. However, this does not happen for the
{\it sine}--definition of the $D_T$ propagator. We associate this
result with insufficient number of the Gribov copies used in the
gauge fixing. The $D_T$ propagator requires many more Gribov
copies for the gauge fixing than does the $D_L$ propagator.

Concerning the other parameters of the fits, the mass extracted
from the propagator at zero temperature and from the $D_L$
propagator at non--zero temperature does not depend on the definition
of the propagator. The mass for the $T=0$ case is perfectly described
by the Polyakov formula. The $m_L$ mass vanishes at the phase
transition point as was expected from the disappearance of the
monopole plasma at the critical temperature. Beyond the phase transition
point, the $m_T$ mass measured in this paper does not behave in a
physical way due to the severe Gribov copy problem.

Finally let us comment on the continuum limit of the measured
quantities. The continuum limit of the cQED$_3$ corresponds to
$\beta= 1/(g_3^2~a) \to \infty$ holding the dimensionful gauge
coupling $g_3$ fixed. According to Polyakov, non-perturbative
quantities such as the Debye mass and string tension can be expressed in
terms of $g_3$ and the monopole density $\rho$, which generally
might be independent quantities. However, this is not true for
compact $U(1)$ where both $g_3$ and $\rho$ depend on a single
parameter, the lattice coupling $\beta$. Therefore, in the limit
of vanishing lattice spacing the monopole density and other
non-perturbative quantities such as the Debye mass, string tension and
anomalous dimension also vanish exponentially as $\sim \exp\{-
{\mathrm{const}} \,\,\beta\}$ ($cf.$
Eqs.~(\ref{def:mass_theor},\ref{def:density_theor})). However, in
more realistic models (like the Georgi-Glashow model) the monopole
density and the lattice spacings are indeed independent and the
monopole density should survive in the continuum limit. According
to our results, this implies that in the continuum limit of such
theories a non--zero anomalous dimension in the photon propagator
can be expected.

\section*{Acknowledgements}

M.~N.~Ch. is supported by the JSPS Fellowship P01023. E.-M.~I.
gratefully appreciates the support by the Ministry of Education,
Culture and Science of Japan (Monbu-Kagaku-sho) and the
hospitality extended to him by H. Toki at the RCNP of Osaka
University.


\begin{thebibliography}{99}

\bibitem{Polyakov}
A.~M.~Polyakov,
Nucl.~Phys.~{\bf B 120}, 429 (1977).

\bibitem{ChSB}
H.~R.~Fiebig and R.~M.~Woloshyn,
Phys.~Rev.~{\bf D 42}, 3520 (1990).

\bibitem{Josephson}
Y.~Hosotani,
Phys.~Lett.~{\bf B 69}, 499 (1977);
V.~K.~Onemli, M.~Tas and B.~Tekin,
JHEP {\bf 0108}, 046 (2001).

\bibitem{HighTc}
G. Baskaran and P.~W.~Anderson,
Phys.~Rev.~{\bf B 37}, 580 (1998);
L.~B.~Ioffe and A.~I.~Larkin,
{\it ibid.}~{\bf B 39}, 8988 (1989);
P.~A.~Lee,
Phys.~Rev.~Lett.~{\bf 63}, 680 (1989);
T.~R.~Morris,
Phys.~Rev.~{\bf D 53}, 7250 (1996).

\bibitem{CISPaper1}
M.~N.~Chernodub, E.--M.~Ilgenfritz and A.~Schiller,
Phys.~Rev.~{\bf D 64}, 054507 (2001).

\bibitem{CISPaper2}
M.~N.~Chernodub, E.--M.~Ilgenfritz and A.~Schiller,
Phys.~Rev.~{\bf D 64}, 114502 (2001).

\bibitem{CISLetter}
M.~N.~Chernodub, E.--M.~Ilgenfritz and A.~Schiller,
Phys.~Rev.~Lett.~{\bf 88}, 231601 (2002).

\bibitem{AnomalousMatter}
H.~Kleinert, F.~S.~Nogueira and A.~Sudb\o \,
Phys.~Rev.~Lett.~{\bf 88}, 232001 (2002);
A. Sudb\o \  {\it et al.}, ibid. {\bf 89}, 226403 (2002).

\bibitem{MatterFields}
M.~N.~Chernodub, E.--M.~Ilgenfritz and A.~Schiller,
Phys. Lett. B {\bf 547}, 269 (2002).

\bibitem{PoZuYe}
M.~I.~Polikarpov, Ken~Yee, and M.~A.~Zubkov,
Phys.~Rev.~{\bf D~48}, 3377 (1993).

\bibitem{PhMon}
R.~J.~Wensley and J.~D.~Stack,
Phys.~Rev.~Lett.~{\bf 63}, 1764 (1989).

\bibitem{DGT}
T.~DeGrand and D.~Toussaint,
Phys.~Rev.~{\bf D 22}, 2478 (1980).

\bibitem{DamgaardHeller}
P.~H.~Damgaard and U.~M.~Heller,
Nucl.~Phys.~{\bf B 309}, 625 (1988).

\bibitem{Bogolubsky:1999}
V.~K.~Mitrjushkin,
Phys.~Lett.~{\bf B 390}, 293 (1997);
I.~L.~Bogolubsky, V.~K.~Mitrjushkin, M.~M\"uller-Preussker and P.~Peter,
Phys.~Lett.~{\bf B 458}, 102 (1999).

\bibitem{CurrentQCD}
P.~Marenzoni, G.~Martinelli, N.~Stella and M.~Testa,
Phys.~Lett.~{\bf B 318}, 511 (1993);
%
P.~Marenzoni, G.~Martinelli and N.~Stella,
Nucl.~Phys.~{\bf B 455}, 339 (1995);
%
D.~B.~Leinweber, J.~I.~Skullerud, A.~G.~Williams and C.~Parrinello,
Phys.~Rev.~{\bf D 60}, 094507 (1999)
[Erratum-ibid.~{\bf D 61}, 079901 (1999)];
%
A.~G.~Williams,
{\it Proc. of 3rd Int. Conf. in Quark Confinement and Hadron Spectrum },
hep-ph/9809201.

\bibitem{Ma}
J.~P.~Ma,
Mod.~Phys.~Lett.~{\bf A 15}, 229 (2000).

\bibitem{Kerler}
W.~Kerler, C.~Rebbi and A.~Weber,
Phys.~Lett.~{\bf B 348}, 565 (1995).

\end{thebibliography}
\end{document}